\documentclass[draftcls,12pt,onecolumn]{IEEEtran}

\usepackage{bm} 
\usepackage{geometry}
\usepackage{fancyhdr} 

\usepackage{amsfonts}
\usepackage{times}
\usepackage[pdftex]{graphicx}
\usepackage{latexsym}
\usepackage{dsfont}
\usepackage{amssymb}
\usepackage{amsmath}
\usepackage{cite}
\usepackage{verbatim}
\usepackage{subfigure}

\newcommand{\figref}[1]{{Fig.}~\ref{#1}} 
\newcommand{\tabref}[1]{{Table}~\ref{#1}}

\newcommand{\Ns}{N_s}

\newcommand{\tr}{\text{tr}}

\def\bb0{{\mathbb{0}}}

\def\ba{{\mathbf{a}}}
\def\bb{{\mathbf{b}}}

\def\bd{{\mathbf{d}}}
\def\bee{{\mathbf{e}}}

\def\bg{{\mathbf{g}}}

\def\bq{{\mathbf{q}}}

\def\bs{{\mathbf{s}}}

\def\by{{\mathbf{y}}}
\def\bz{{\mathbf{z}}}
\def\b0{{\mathbf{0}}}

\def\bA{{\mathbf{A}}}
\def\bB{{\mathbf{B}}}
\def\bC{{\mathbf{C}}}
\def\bD{{\mathbf{D}}}

\def\bF{{\mathbf{F}}}

\def\bH{{\mathbf{H}}}
\def\bI{{\mathbf{I}}}
\def\bJ{{\mathbf{J}}}

\def\bW{{\mathbf{W}}}

\def\bbC{{\mathbb{C}}}

\def\bbE{{\mathbb{E}}}

\def\bbR{{\mathbb{R}}}

\def\cA{\mathcal{A}}

\def\cC{\mathcal{C}}
\def\cD{\mathcal{D}}

\def\cH{\mathcal{H}}
\def\cI{\mathcal{I}}

\def\cN{\mathcal{N}}
\def\cO{\mathcal{O}}

\def\cX{\mathcal{X}}
\def\cY{\mathcal{Y}}

\def\sf0{{\mathsf{0}}}

\usepackage{color,graphicx,epsfig,psfrag}
\usepackage{pifont,enumerate,cases,algorithm,balance}
\usepackage{mathrsfs} 
\usepackage{multirow,multicol}
\usepackage[table]{xcolor}

\newcommand{\sref}[1]{{Section}~\ref{#1}} 
\newcommand{\algoref}[1]{{Algortihm}~\ref{#1}}

\def\b0{\mathbf 0}
\def\b1{\mathbf 1}
\def\teq{\triangleq}
\def\new{\text{new}}
\def\old{\text{old}}
\def\j{\text{j}}
\def\t{\text{t}}
\def\r{\text{r}} 
\def\T{\text{T}}
\def\R{\text{R}}
\def\I{\text{I}}
\def\bHd{{\bm{H}_d}}
\def\bAR{{\bA_{\text{R}}}}

\def\bAT{{\bA_{\text{T}}}}
\def\bATH{{\bA_{\text{T}}^*}}

\def\bARv{{\bA^{\text{v}}_{\text{R}}}}
\def\bATv{{\bA^{\text{v}}_{\text{T}}}}
\def\baR{{\ba_{\text{R}}}}

\def\baT{{\ba_{\text{T}}}}

\def\olbA{{\overline{\bA}}}
\def\olbAT{{\overline{\bA}_{\text{T}}}}
\def\olbDT{{\overline{\bD}_{\text{T}}}}
\def\olbCT{{\overline{\bC}_{\text{T}}}}
\def\olbGammaT{{\overline{\bm \Gamma}_{\text{T}}}}
\def\bDR{{\bD_{\text{R}}}}
\def\bDT{{\bD_{\text{T}}}}
\def\bDTH{{\bD^*_{\text{T}}}}

\def\bmu{{\bm \mu}}
\def\bmuw{{\bm \mu_{\text{w}}}}
\def\bxi{{\bm \xi}}
\def\btheta{{\bm \theta}}
\def\bphi{{\bm \phi}}

\def\bOmegad{{\mathbf\Omega_d}}
\def\bOmegac{{\mathbf\Omega[c]}}
\def\bDeltad{{\mathbf \Delta_d}}
\def\bDeltadv{{\mathbf \Delta_d^{\text{v}}}}
\def\bDeltac{{\mathbf \Delta[c]}}
\def\prc{{p_{\text{rc}}}}

\def\thetalk{{\theta_{\ell,k}}}
\def\phil{{\phi_{\ell}}}
\def\philk{{\phi_{\ell,k}}}
\def\taul{{\tau_{\ell}}}
\def\bPhi{{\mathbf{\Phi}}}
\def\bPhiw{{\mathbf{\Phi}_{\text{w}}}}
\def\bPsi{\mathbf{\Psi}}

\def\bGammaR{\mathbf{\Gamma}_{\R}}
\def\bGammaT{\mathbf{\Gamma}_{\T}}
\def\bGammaTH{\mathbf{\Gamma}^*_{\T}}
\def\bCR{\mathbf{C}_{\R}}
\def\bCT{\mathbf{C}_{\T}}
\def\bCTH{\mathbf{C}^*_{\T}}

\def\wbG{\widetilde{\mathbf G}}
\def\wbh{\widetilde{\mathbf h}}
\def\wbhu{\widetilde{\mathbf h}^{(u)}}
\def\wbH{\widetilde{\mathbf H}}
\def\wbHu{\widetilde{\mathbf H}^{(u)}}
\def\wbX{\widetilde{\mathbf X}}
\def\wby{\widetilde{\mathbf y}}
\def\wbY{\widetilde{\mathbf Y}}
\def\wbz{\widetilde{\mathbf z}}
\def\wbZ{\widetilde{\mathbf Z}}

\def\hbH{\hat{\mathbf H}}
\def\hbU{\hat{\mathbf U}}

\def\Nt{{N_{\text{t}}}}
\def\Nr{{N_{\text{r}}}} 
\def\Gt{{G_{\text{t}}}}
\def\Gr{{G_{\text{r}}}}
\def\Ns{{N_{\text{s}}}}  
\def\Lt{{L_{\text{t}}}}
\def\Lr{{L_{\text{r}}}}
\def\Kt{{K_{\text{t}}}}
\def\Kr{{K_{\text{r}}}}
\def\Nc{{N_{\text{c}}}}
\def\Nsa{{N_{\text{sa}}}}
\def\Np{{N_{\text{p}}}}
\def\Nray{{N_{\text{ray}}}}
\def\Nrepeat{{N_{\text{rep}}}}
\def\NQ{{N_{\text{Q}}}}
\def\Ntap{{N_{\text{tap}}}}
\def\Ts{{T_{\text{s}}}} 
\def\BB{{\text{BB}}}
\def\RF{{\text{RF}}}
 
\def\vec{\mathrm{vec}}
\def\supp{\text{supp}}
\def\kron{\otimes} 
\def\diag{\text{diag}}

\usepackage{bm} 
\def\bcA{\bm{\cA}}

\def\bcH{\bm{\cH}}

\def\bcX{\bm{\cX}}
\def\bcY{\bm{\cY}}

\def\bcHu{\bm{\mathcal{H}}^{(u)}}

\def\bcXu{\bm{\mathcal{X}}^{(u)}}

\def\bsfH{\bm{\mathsf{H}}}

\def\bsfX{\bm{\mathsf{X}}}
\def\bsfY{\bm{\mathsf{Y}}}
\def\bsfZ{\bm{\mathsf{Z}}}

\begin{document}
\title{Dictionary Learning for Channel Estimation in Hybrid Frequency-Selective mmWave MIMO Systems}
\author{Hongxiang Xie, \emph{Student Member, IEEE}, Javier Rodr\'{i}guez-Fern\'{a}ndez, \emph{Student Member, IEEE}, and Nuria Gonz\'{a}lez-Prelcic, \emph{Senior Member, IEEE.}
\thanks{H. Xie, J. Rodr\'{i}guez-Fern\'{a}ndez, and N. Gonz\'{a}lez-Prelcic are with The University of
Texas at Austin, Austin, TX 78701 USA (e-mail: \{xiehx, javi.rf, ngprelcic\}@utexas.edu).
N. Gonz\'{a}lez-Prelcic is also with the University of Vigo, Spain.
This material is based upon work supported in part by the National Science Foundation under Grant No. CNS-1702800, and the Spanish Government and the European Regional Development Fund (ERDF) under project MYRADA (TEC2016-75103-C2-2-R).} 
} 
\maketitle 
\thispagestyle{empty}

\begin{abstract}  
Exploiting channel sparsity at millimeter wave (mmWave) frequencies reduces the high training overhead associated with the channel estimation stage. Compressive sensing (CS) channel estimation techniques usually adopt the (overcomplete) wavelet/Fourier transform matrix as a sparsifying dictionary. This may not be the best choice when considering non-uniform arrays, antenna gain/phase errors, mutual coupling effects, etc. We propose two dictionary learning (DL) algorithms  to learn the best sparsifying dictionaries for channel matrices from observations obtained with hybrid frequency-selective mmWave multiple-input-multiple-output (MIMO) systems. First, we optimize the combined dictionary, i.e., the Kronecker product of transmit and receive dictionaries, as it is used in practice to sparsify the channel matrix. Second, considering the different array structures at the transmitter and receiver, we exploit separable DL to find the best transmit and receive dictionaries. Once the channel is expressed in terms of the optimized dictionaries, various CS-based sparse recovery techniques can be applied for low overhead channel estimation. The proposed DL algorithms perform well under low SNR conditions inherent to any mmWave communication systems before the precoders/combiners can be optimized. The effectiveness of the proposed DL algorithms has been corroborated via numerical simulations with different system configurations, array geometries and hardware impairments.

\end{abstract} 


\section{Introduction}\label{sec:intro} 
To reduce the high overhead associated to estimating the channel in  mmWave MIMO systems,   channel spatial sparsity has been exploited, e.g.,  \cite{HanLee:Two-stage-compressed-sensing:16, LeeGilLee:Exploiting-spatial-sparsity:14}. In most prior work, a narrowband channel model is considered, and the sparse channel matrices under a certain sparsifying dictionary are recovered from compressive channel measurements with few training resources.  The dictionaries used in prior work are constructed from the transmit and receive array response vectors evaluated on a grid of quantized possible angles of arrival and departure (AOAs/AODs) \cite{HeaGonRan:An-Overview-of-Signal-Processing:16,RodGonVen:Frequency-domain-Compressive-Channel:18}.
The sparse nature of frequency selective mmWave MIMO channels, both in the angular and delay domains, has also been considered to redefine the sparsifying dictionaries \cite{VenAlkPre:Channel-Estimation-for-Hybrid:17}. 
Unfortunately, in prior work \cite{HanLee:Two-stage-compressed-sensing:16, LeeGilLee:Exploiting-spatial-sparsity:14,HeaGonRan:An-Overview-of-Signal-Processing:16,RodGonVen:Frequency-domain-Compressive-Channel:18,VenAlkPre:Channel-Estimation-for-Hybrid:17}, perfect antenna array manifolds in the channel model, without taking into consideration many practical effects, including hardware impairments, calibration errors and so on. 

Practically constructed antenna arrays deviate from the ideal case in many ways. Due to the manufacture and calibration errors, the antenna array will generate unexpected radiation patterns (including amplitude and phase patterns). 
The imperfect spacing between antenna elements also have to be considered. For example, the antenna spacing between uniform linear array (ULA) elements in practical arrays is not the ideal half-wavelength due to the limited manufacturing accuracy, which will result in  irregular linear arrays rather than perfect ULAs. Therefore, the array response vectors do not longer follow the Vandermonde structure. The disturbance in the antenna spacing further induces the mutual coupling effect among antenna elements. There are also other hardware impairments in the radio frequency (RF) chains and the calibration errors contribute to a general mismatch between ideal and actual channel models.  Given all the sources of mismatch, sparsifying dictionaries constructed from ideal array response vectors at quantized angels are no longer the best choice for exploiting channel sparsity.  

Learning a sparsifying dictionary using DL is one approach to capture the underlying practical structure in mmWave MIMO channels. This way, the compressive channel estimation stage will have the capability to adapt to all kinds of uncertainties and impairments.  
DL for sparse signal representation has many applications in image processing including image denoising  \cite{ElaAha:Image-Denoising-Via-Sparse:06,ZhaElyAer:Novel-Methods-for-Multilinear:14}, component analysis \cite{BahPanZaf:Robust-Kronecker-Component:18}, classification \cite{MaiPonSap:Supervised-Dictionary-Learning:09}, or feature extraction  \cite{NieHuaCai:Efficient-and-Robust-Feature:10}, and among others \cite{LiuCheChe:Support-Discrimination-Dictionary:16,HawSeiKle:Separable-dictionary-learning:13}. 
DL for wireless signal processing is not straightforward, however, given the different signal characteristics, signal-to-noise (SNR) operation ranges and sparsity structures. The idea of DL-based channel sparse representation and estimation was proposed in \cite{DinRao:Dictionary-Learning-Based-Sparse:18} for massive MIMO systems operating at low frequencies. In that work, an overcomplete dictionary was learned from the channel measurement (training) data to substitute the predetermined discrete Fourier transform (DFT) dictionaries. 
A similar approach for DL-based low-rank channel approximation was also considered in \cite{WieWeiUts:Low-Rank-Approximations-for-Spatial:16}.
While \cite{DinRao:Dictionary-Learning-Based-Sparse:18} shows the power of leveraging DL for channel state information (CSI) acquisition, the formulation was limited to narrow band
massive MIMO systems in high SNR regimes, without considering the aforementioned practical
effects or the operating conditions at mmWave frequencies. 

In this paper, we develop DL strategies for frequency selective mmWave MIMO systems with hybrid array architectures. The main contributions of this paper are as follows: 
\begin{itemize} 
	\item We propose a general model for frequency-selective mmWave MIMO systems that explicitly includes the array manifold disturbances, antenna gain/phase errors, array mutual coupling, and so on. This general model will motivate and justify the formulation of our DL problems. 
	\item We propose the \textit{combined dictionary learning (CoDL)} algorithm to directly optimize a combined dictionary, i.e., the Kronecker product of transmit and receive dictionaries. By exploiting the common sparsity between subcarriers, CoDL is formulated as a non-convex optimization problem with two regularization terms to promote the common sparsity and combat high noise level. 
	While common sparsity has been commonly considered for CS-based channel estimation, e.g., \cite{RodGonVen:Frequency-domain-Compressive-Channel:18,GaoDaiDai:Structured-Compressive-Sensing-Based:16,GaoDaiWan:Spatially-Common-Sparsity:15}, it has never been exploited for learning a sparsifying dictionary for wireless channels. 
	\item We propose the \textit{separable dictionary learning (SeDL)} algorithm to optimize the transmit and receive dictionaries separately, which is more consistent with the practical system architecture, considering different array structures at transmitters and receivers.
	To exploit the separability of the Kronecker product between transmit and receive dictionaries, we formulate the SeDL problem in the tensor space, where the common sparsity among subcarriers is translated into the common sparsity support along the third-dimension of tensors. This is a typical property and constraint for our SeDL formulation that has not been considered in existing CS or DL problems.   
	Though there is a performance gap between SeDL and CoDL, SeDL has a much lower computational complexity due to the smaller sizes of transmit/receive dictionaries compared to the combined one used in CoDL. Therefore, SeDL achieves a good trade-off between performances and complexity. 
	\item We derive the Cram\'{e}r-Rao Lower Bound (CRLB) for the estimation variance of frequency-domain channel matrices with unknown dictionaries. This helps to understand the performances of various compressive channel estimation techniques with different sparsifying dictionaries. 
	\item We evaluate the proposed DL algorithms on different system configurations, array geometries and channel conditions. Numerical results show that the training overhead of channel estimation with learned dictionaries can be significantly reduced compared to that based on overcomplete dictionaries constructed from array response vectors. This corroborates the effectiveness of the proposed DL algorithms for hybrid wideband mmWave MIMO systems.   
\end{itemize} 
Compared with our prior work in \cite{XieGonHea:Seprable-dictionary-learning:19}, we developed new algorithm, CoDL, and compared with the CRLB. 
The rest of the paper is organized as follows. \sref{sec:systemmodel} describes the system, channel and signal models for the considered wideband mmWave MIMO system based on fully connected hybrid  architectures.
The CRLB computation for the problem of estimating a mmWave channel with unknown dictionaries is described in \sref{sec:CRLB}.
\sref{sec:CoDL} and \sref{sec:SeDL} introduce the proposed CoDL and SeDL algorithms, respectively. Numerical simulations are provided in \sref{sec:Simulations} to justify the effectiveness of the proposed DL algorithms and conclusions are drawn in \sref{sec:conclusions}. 

\textbf{Notations:} Vectors and matrices are denoted by boldface small and
capital letters;  the transpose, conjugate, Hermitian (conjugate transpose),
inverse, and pseudo-inverse of the matrix $\bA$ are denoted by 
$\bA^T$, $\olbA$, $\bA^*$, $\bA^{-1}$ and $\bA^\dag$;  
${\bI_M}$ is an $M\times M$ identity matrix;   
$\mathbf 0_{M\times N}$ is an $M\times N$ all-zero matrix and
$\mathbf 1_{M\times N}$ is an $M\times N$ all-one matrix; 
$[\bA]_{:,j}$ denotes the $j$-th column vector of $\bA$; 
$\teq$ represents new definitions;  
$\cI(N)\teq \{0,1,\ldots,N-1\}$ denotes the index set of cardinality $N$; 
$\bbC$ and $\bbR$ denote the sets of complex and real numbers; 
$\bbE\{\cdot\}$ returns expectation; 
$\text{tr}\{\bA\}$ is the trace of $\bA$; $\lfloor x\rfloor$ denotes the largest integer less than or equal to $x$; $\j = \sqrt{-1}$ is the imaginary unit;
$\kron$,  $\odot$ an $\star$ denote the Kronecker, Hadamard and Khatri-rao product; 
tensors are denoted by bold-faced calligraphic capital 
letters, e.g., $\bcA$. For an $Q$-dimensional ($Q$-order) tensor 
$\bcA\in\bbC^{M_1\times M_2\times \ldots \times M_Q}$, the $q$-mode unfolding, denoted by $[\bcA]_{(q)}\in\bbC^{M_q\times M_1\cdots M_{q-1}M_{q+1}\cdots M_Q}$, represents a rearrangement of $\bcA$ into a matrix, where the $q$-th index is used as a row index and all other indices are aligned along the columns
(aligned in reverse cyclical ordering), and the columns of $[\bcA]_{(q)}$ are referred to as mode-$q$ fibers (columns). The $q$-mode product between a tensor $\bcA\in\bbC^{M_1\times M_2\times \ldots \times M_Q}$ and a matrix
$\bB \in\bbC^{N_q\times M_q}$ is denoted by ${\bcA}_{~\times q~} \bB$ and defined as $[{\bcA}_{~\times q~} \bB]_{(q)} = \bB[\bcA]_{(q)}$. Moreover, there is 
$\bcY=\bcA_{~\times 1~}{\bB^{(1)}}_{\times 2~}{\bB^{(2)}\cdots}_{\times Q~}\bB^{(Q)} \Leftrightarrow [\bcY]_{(q)}=\bB^{(q)}[\bcA]_{(q)}\left(\bB^{(Q)}\kron \cdots \kron \bB^{(q+1)}\kron\bB^{(q-1)}\kron\cdots\kron\bB^{(1)}\right)^T$. 
Furthermore, for a third-order 
tensor $\bcA\in\bbC^{M_1\times M_2 \times M_3 }$, its $i$-th  horizontal, $j$-th lateral, and $c$-th frontal slides are denoted by $[\bcA]_{i::}\in\bbC^{M_2 \times M_3}$, $[\bcA]_{:j:}\in\bbC^{M_1 \times M_3}$,
and $[\bcA]_{::c}\in\bbC^{M_1 \times M_2}$.

\section{System and Channel Models}\label{sec:systemmodel}

\subsection{System model}

Consider a hybrid mmWave multi-user MIMO system with an access point (AP) of $\Nt$ antennas and $\Lt$ RF chains, as well as user equipments (UEs) using $\Nr$ antennas and $\Lr $ RF chains. The channel between the AP and the UE is assumed to be frequency-selective. An orthogonal frequency division multiplexing (OFDM)-based mmWave MIMO link employing $\Nc$ subcarriers is used to simultaneously transmit $\Ns~(\leq \min\{\Lt,\Lr\})$ data streams. 
The hybrid precoder and combiner adopted for such frequency-selective mmWave systems can be represented as 
$\bF[c] = \bF_{\RF}\bF_{\BB}[c]$ 
and $\bW[c] = \bW_{\RF}\bW_{\BB}[c] $, 
for the $c$-th ($c\in\cI(\Nc)$) subcarrier,
where $\bF_{\RF}\in\bbC^{\Nt\times \Lt}$ 
and $\bF_{\BB}[c]\in\bbC^{\Lt\times \Ns}$ 
denote the analog and digital precoders, and $\bW_{\RF}\in\bbC^{\Nr\times \Lr} $ and $\bW_{\BB}[c]\in\bbC^{\Lr\times \Ns}$ are the analog and digital combiners. 
The analog precoders/combiners are frequency-flat, while the baseband ones can be different for each subcarrier. 
In this manuscript, we will consider a fully connected phase shifting network for the analog precoder and combiner.    
During the channel estimation stage, prior knowledge of the training precoders and combiners is assumed at both the AP and the UE. 

\subsection{Channel model} 

We consider the frequency-selective channel model in  \cite{SchSay:Channel-estimation-and-precoder:14,VenAlkPre:Channel-Estimation-for-Hybrid:17}, consisting of $\Np$ clusters with $\Nray$ rays in each cluster and a delay tap length $\Ntap$.  In the sequel, we will focus on DL and channel estimation for the downlink, although the analysis and proposed algorithms can be similarly developed for the uplink. 
The $d$-th delay tap of the downlink channel  between the AP and a UE is denoted as $\bHd\in\bbC^{\Nr \times \Nt}, ~d\in\cI(\Ntap)$ and can be expressed as 
\begin{align}\label{equ:Hd} 
 \bHd =  \sqrt{\frac{\Nt\Nr}{\Np\Nray}}\sum_{\ell=1}^{\Np}\sum_{k=1}^{\Nray} \alpha_{\ell,k} \prc(d\Ts-\taul) \bCR\bGammaR\baR(\philk)(\bCT\bGammaT\baT(\thetalk))^*,  
\end{align} 
where $\prc(\tau)$ denotes a band-limited function including all filtering effects evaluated at $\tau$; $\Ts$ is the system sampling time;
$\alpha_{\ell,k}\in\bbC$ is the complex gain; $\philk\in[-\pi,\pi)$ and $\thetalk\in[-\pi,\pi)$ are the AOA 
and AOD of the $k$-th ray in the $\ell$-th cluster;  $\taul\in \bbR$ is the path delay of all rays in the $\ell$-th cluster.
Moreover, $\baR(\philk)\in\bbC^{\Nr\times 1}$ and $\baT(\thetalk)\in\bbC^{\Nt\times 1}$ denote the antenna array response vectors at UE and AP, which depend on the specific geometries of the antenna arrays and include any disturbance in the spacing between antenna elements due to manufacture errors. For instance, for a linear antenna array, instead of assuming a perfect ULA with an ideal uniform antenna spacing $d$, we denote $\baR(\phi)$ as 
\begin{align} \label{equ:aR}
	\baR(\phi)\teq\frac{1}{\sqrt{\Nr}}\big[1,e^{-\j2\pi (d+\epsilon_{\r,1})/\lambda\sin(\phi)},e^{-\j2\pi (2d+\epsilon_{\r,2})/\lambda\sin(\phi)},\ldots,e^{-\j2\pi ((\Nr-1)d +\epsilon_{\r,\Nr-1})/\lambda\sin(\phi)} \big], 
\end{align} 
where $\lambda$ is the carrier wavelength and $\epsilon_{\r,1},\ldots, \epsilon_{\r,\Nr-1}$ denote the errors in the spacing between receive antenna elements.  
Furthermore, $\bCR\in\bbC^{\Nr\times \Nr}$ and $\bCT\in\bbC^{\Nt\times \Nt}$ in \eqref{equ:Hd} are the mutual coupling matrices for the receive and transmit antenna arrays, representing the unwanted interchange of energy between elements in the arrays \cite{EbeEscBie:Investigations-on-antenna-array:16}.
$\bGammaR\in\bbC^{\Nr\times \Nr}$ and $\bGammaT\in\bbC^{\Nt\times \Nt}$ are the antenna gain and phase error matrices, defined as $\bGammaR\teq\diag\big\{g_{\r,1}e^{\j\nu_{\r,1}},g_{\r,2}e^{\j\nu_{\r,2}},\ldots,g_{\r,\Nr}e^{\j\nu_{\r,\Nr}}\big\}$, in which $\{g_{\r,i}\}_{i=1}^{\Nr}$ are the receive gain error normalized to a reference amplitude and $\{\nu_{\r,i}\}_{i=1}^{\Nr}$ are the additional receive phase errors.
Note that these antenna gain and phase errors are due to the hardware impairments and calibration errors in production processes with respect to impedance matching networks, baluns, possible amplifiers, PCB materials, etc, \cite{EbeEscBie:Investigations-on-antenna-array:16}.

We define an $\Np\Nray
{\times} \Np\Nray$ diagonal matrix that contains the channel coeffients as $\bDeltad \teq  \sqrt{\frac{\Nt\Nr}{\Np\Nray}}~\diag\big\{\alpha_{1,1} \prc(d\Ts-\tau_1),\ldots, \alpha_{\Np,\Nray} \prc(d\Ts-\tau_{\Np})\big\}$. Then the compact expression for \eqref{equ:Hd} is given as 
\begin{align}\label{equ:HdArrdict}
	\bHd = \bCR\bGammaR\bAR \bDeltad \bATH\bGammaTH\bCTH, 
\end{align}
where $\bAR\teq \big[\baR(\phi_{1,1}),\ldots, \baR(\phi_{\Np,\Nray}) \big]$ 
and $\bAT\teq \big[\baT(\theta_{1,1}),\ldots, \baT(\theta_{\Np,\Nray}) \big]$
collect the receive and transmit array response vectors evaluated at the actual AOAs and AODs. 

To exploit the sparsity within the channel matrix and enable the CS techniques, the exact expression of $\bHd$ in \eqref{equ:HdArrdict}  
can be approximated with the extended virtual channel model \cite{ HeaGonRan:An-Overview-of-Signal-Processing:16} as 
\begin{align}\label{equ:HdVCM} 
	\bHd \approx \bCR\bGammaR\bARv \bDeltadv (\bATv)^*\bGammaTH\bCTH, 
\end{align} 
where the dictionary matrices $\bARv\in\bbC^{\Nr\times \Gr}$ and $\bATv\in\bbC^{\Nt\times \Gt}$ generalize $\bAR$ and $\bAT$ in \eqref{equ:HdArrdict}, 
while $\bDeltadv\in\bbC^{\Gr\times \Gt}$ is the generalization of $\bDeltad$ in \eqref{equ:HdArrdict}. 
$\bARv$ and $\bATv$ collect the receive and transmit array response vectors evaluated on $\Gr$ quantized angles for AOAs and $\Gt$ quantized angles for AODs, both sampled in $[-\pi,\pi)$, and $\bDeltadv$ contains the path gains of these discrete quantized AOAs/AODs at the non-zero elements.   
Inspecting \eqref{equ:HdVCM}, if there is no prior knowledge on $\bCR$, $\bCT$, $\bGammaR$, $\bGammaT$, the existing channel estimation strategies based on CS techniques will not be applicable for this general model as the dictionary would be unknown. For this reason,
in prior work like  \cite{VenAlkPre:Channel-Estimation-for-Hybrid:17,RodGonVen:Frequency-domain-Compressive-Channel:18}, the mutual coupling matrices and gain/phase error matrices were all set as identity matrices and the antenna spacing disturbances were considered as zeros. Under this circumstance, a popular choice for $\bARv$ and $\bATv$ is the overcomplete DFT matrices if perfect ULAs are considered at the AP and the UE. Nevertheless, this is not an optimal choice for the general channel models that include hardware imperfections and callibration errors. 

A natural solution is to substitute $\bCR\bGammaR\bARv$ and $\bCT\bGammaT\bATv$ in \eqref{equ:HdVCM} with two general dictionaries $\bDR$ and $\bDT$, without any array structure related constraints, so that they can be applied to
arbitrary antenna geometries and include all the hardware impairments. Under these assumptions, 
$\bHd$ in \eqref{equ:HdArrdict}  
can be generalized as
\begin{align}\label{equ:HdDL}        
	\bHd \approx \bDR \bOmegad \bDTH,
\end{align}
where $\bDR\teq[\bd_{\R,0},\bd_{\R,1},\ldots,\bd_{\R,\Kr-1}]\in\bbC^{\Nr\times \Kr}$ and
$\bDT\teq[\bd_{\T,0},\bd_{\T,1},\ldots,\bd_{\T,\Kt-1}]\in\bbC^{\Nt\times \Kt}$
denote the optimal receive and transmit dictionaries to be determined, and 
$\bOmegad\in\bbC^{\Kr\times \Kt}$ is a sparse channel matrix with few non-zero elements, similar to its counterpart $\bDeltadv$ in \eqref{equ:HdVCM}. 
To avoid the ambiguity between dictionaries and channel matrices, the dictionary atoms (columns) are normalized, i.e., $\|\bd_{\R,k_r}\|_2=1, \forall k_r\in\cI(\Kr)$ and $\|\bd_{\T,k_t}\|_2=1, \forall k_t\in\cI(\Kt)$, where $\Kr ~(\geq \Nr)$ and $\Kt ~(\geq \Nt)$ are the numbers of atoms of each dictionary. 
Note that with the optimized dictionaries accounting for practical antenna uncertainties and adapted to different channel effects, it is expected that the new channel matrix $\bOmegad$ will be sparser than $\bDeltadv$. 

For the geometric channel model in \eqref{equ:HdDL}, the frequency-domain channel matrix at the $c$-th ($c\in\cI(\Nc)$) subcarrier can be written as 
\begin{align}\label{equ:Hk} 
	\bH[c] &= \sum_{d=0}^{\Ntap-1}\bHd e^{-\j\frac{2\pi c d}{\Nc}}  
	 = \bCR\bGammaR\bAR \underbrace{\bigg(\sum_{d=0}^{\Ntap-1} \bDeltad e^{-\j\frac{2\pi c d}{\Nc}} \bigg)}_{\bDeltac}\bATH\bGammaTH\bCTH   \notag\\ 
	 & \approx \bDR\underbrace{\bigg(\sum_{d=0}^{\Ntap-1} \bOmegad e^{-\j\frac{2\pi c d}{\Nc}} \bigg)}_{\bOmegac}\bDTH,
\end{align} 
where $\bDeltac$ and $\bOmegac$ are defined accordingly and denote the channel gains in the frequency domain. To be noted, we assume that $\bCR,\bCT,\bGammaR,\bGammaT$ as well as $\bAR$ and $\bAT$ are frequency-independent in this paper, as we neglect the beam squint effect \cite{WanGaoJin:Spatial--and-Frequency-Wideband-Effects:18,BraSay:Wideband-communication-with:15}. Therefore, the generalized dictionaries $\bDR$ and $\bDT$ are also frequency-independent in \eqref{equ:Hk}. 
Recalling that $\vec(\bA\bB\bC) = ({\bC}^T\kron\bA)\vec(\bB)$, the vectorization of \eqref{equ:Hk} is given as   
\begin{align}\label{equ:Hkvect}    
	\vec\big(\bH[c]\big)  
	=\big((\olbCT\olbGammaT\olbAT)\kron (\bCR\bGammaR\bAR)\big)
	\vec\big(\bDeltac\big) 
	\approx \big(\olbDT \kron \bDR\big)
	\vec\big(\bOmegac\big) 
	= \bPsi\wbh[c], 
\end{align} 
where $\bPsi \teq \big(\olbDT \kron \bDR\big)\in\bbC^{\Nr \Nt \times \Kr \Kt}$ is the combined dictionary and 
$\wbh[c]\teq \vec(\bOmegac)\in\bbC^{\Kr \Kt\times 1}$ is the vectorized sparse channel matrix when this combined dictionary is used to build $\vec{(\bH[c])}$.  

\subsection{Frequency domain signal model}

According to  \cite{VenAlkPre:Channel-Estimation-for-Hybrid:17,RodGonVen:Frequency-domain-Compressive-Channel:18}, the received signal at the UE for the $c$-th subcarrier can be written as  
\begin{align}\label{equ:yc} 
	\by[c] = \bW_{\BB}^*[c]\bW_{\RF}^* \bH[c]\bF_{\RF}\bF_{\BB}[c]\bs[c] 
	+ \bW_{\BB}^*[c]\bW_{\RF}^* \bz[c],  
\end{align} 
where $\bs[c]\in\bbC^{\Ns\times 1}$ is the transmitted signal vector at the $c$-th subcarrier and $\bz[c]\sim \cC\cN(\mathbf 0,\sigma^2\bI_{\Nr})$
denotes the Gaussian noise vector with variance  $\sigma^2$.   
During the DL and channel estimation phases, let the AP and the UE use the same frequency-flat precoder $\bF_m\in\bbC^{\Nt\times \Lt}$ and 
combiner $\bW_m\in\bbC^{\Nr\times \Lr}$  for all the subcarriers in the $m$-th OFDM symbol.
Suppose that the transmitted symbols satisfy $\mathbb{E}\big\{\bs_m[c]\bs_m^*[c] \big\}=\frac{P_{\text{tr}}}{\Ns}\bI_{\Ns}$, with 
$P_{\text{tr}}$ the total power constraint and thus the SNR is defined as $\text{SNR}=\frac{P_{\text{tr}}}{\sigma^2}$.  
To facilitate our proposed DL and channel estimation algorithms in the following, we decompose the transmitted symbol as $\bs_m[c] = \bq_mt_m[c]$ with 
$\bq_m\in\bbC^{\Lt\times 1}$ a frequency-flat vector  
and $t_m[c]$ a scalar pilot symbol known at the receiver, as in  \cite{RodGonVen:Frequency-domain-Compressive-Channel:18}.  
To provide SNR gain for the DL algorithm to work properly, additional temporal spreading has to be considered. Therefore, during the learning stage, the training sequence is generated by repetition of the symbols in the auxiliary symbol sequence $r_n[c], n=0,1,2,\ldots$ For a given spreading factor $\Nrepeat$, the training sequence is  generated as $t_m[c]=r_{\lfloor\frac{m}{\Nrepeat}\rfloor}$[c], i.e.,
\begin{align}
t_m[c]=\underbrace{r_0[c],\cdots,r_0[c]}_{\Nrepeat~\text{times}},\underbrace{r_1[c],\cdots,r_1[c]}_{\Nrepeat~\text{times}},\cdots 
\end{align} 
Then the post-combining received training signals at $c$-th subcarrier, i.e., \eqref{equ:yc}, in the $m$-th ($m=0,1,2,\ldots$) training OFDM symbol is rewritten as
\begin{align} 
\by_m[c] = \bW_m^*\bH[c] \bF_m\bq_m t_m[c] + \bW_m^*\bz_m[c]. 
\end{align}
To enable sparse reconstruction with a single subcarrier-independent measurement matrix, we multiply the received signal $\by_m[c]$ by $(t_m[c])^{-1}$ and vectorize it to get
\begin{align}\label{equ:ym}
\wby_m[c] \teq\vec\big((t_m[c])^{-1}\by_m[c]\big) =\left(\bq_m^T\bF_m^T\kron\bW_m^*\right)\vec\left(\bH[c]\right) + \wbz_m[c],
\end{align} 
where $\wbz_m[c]\teq \vec((t_m[c])^{-1}\bW_m^*\bz_m[c])$. 
Combining with \eqref{equ:Hkvect}, \eqref{equ:ym} can be approximated by 
\begin{align}\label{equ:DLexpr}
\wby_m[c] &\approx\left(\bq_m^T\bF_m^T\kron\bW_m^*\right)\big(\olbDT \kron \bDR\big)
\vec\big(\bOmegac\big) + \wbz_m[c] 
 =\bPhi_m\bPsi \wbh[c]+ \wbz_m[c], 
\end{align} 
where 
$\label{equ:Phi} 
\bPhi_m \teq \left(\bq_m^T\bF_m^T\kron\bW_m^*\right) \in\bbC^{\Lr\times \Nr \Nt}
$ is defined accordingly. Note that $\bPhi_m$ is the sensing matrix for $m$-th OFDM symbol based on hybrid precoders and combiners, and $\bPhi_m \bPsi$ is the measurement matrix commonly defined in the literature of CS. To use the training spreading to average out the noise, we will keep the same $\bF_m$ and $\bW_m$ for the $\Nrepeat$ OFDM symbols during which the same training symbol $t_m[c]$ is transmitted, and then averaged the received signals as $\wby_{\text{ave},i}[c]=\sum_{m=0}^{\Nrepeat-1}\wby_{i\Nrepeat+m}[c]/\Nrepeat$ for $i=0,1,2,\ldots$.  Next, we formulate a tall sensing matrix by stacking $M$ averaged measurements obtained above, and express the received signals at the $c$-th subcarrier as
\begin{align}\label{equ:Mframe}
\underbrace{\Big[\wby_{\text{ave},0}^T[c],  \ldots, \wby_{\text{ave},M-1}^T[c] \Big]^T}_{\wby[c]} &= 
\underbrace{\Big[\bPhi_0^T, \ldots, \bPhi_{M-1}^T\Big]^T}_{\bPhi}	
\bPsi\wbh[c] + 
\underbrace{\Big[\wbz_{\text{ave},0}^T[c],\ldots,\wbz_{\text{ave},M-1}^T[c]\Big]^T}_{\wbz[c]},
\end{align}
where $\wby[c]\in\bbC^{M\Lr \times 1}$, $\bPhi\in\bbC^{M\Lr \times \Nr \Nt}$ and $\wbz[c]\in\bbC^{M\Lr \times 1}$ are defined accordingly.

\section{ CRLB analysis for channel estimation with unknown dictionaries} 
\label{sec:CRLB} 

In this section, we compute the CRLB \cite{Kay:Fundamentals-of-statistical-signal:93, RodGonVen:Frequency-domain-Compressive-Channel:18} on the variance of unbiased estimators of the frequency-domain channel matrix $\bH[c], c\in\cI(\Nc)$, without assuming prior knowledge on the dictionaries. Specifically, from \eqref{equ:HdArrdict} and \eqref{equ:Hk}, we first rewrite the vectorized channel matrix as $\vec(\bH[c])=\big((\olbCT\olbGammaT\olbAT(\btheta))\star \bCR\bGammaR\bAR(\bphi)\big)\bg[c]$ and $\bg[c]\in\bbC^{\Np\Nray \times 1}$ contains the nonzero diagonal elements of $\bDeltac$ corresponding to the multipath gains at the $c$-th subcarrier. 
For the estimation of the channel matrices $\bH[c], c\in\cI(\Nc)$, we collect the measurements at all $\Nc$ subcarriers based on \eqref{equ:Mframe}, and rewrite it as 
	\begin{align}  
	\underbrace{\big[\wby[0],\ldots,\wby[\Nc-1]\big]}_{\wbY}
	&=\bPhi \bPsi\underbrace{\big[\wbh[0],\ldots,\wbh[\Nc-1]\big]}_{\wbH} +\underbrace{\big[\wbz[0],\ldots,\wbz[\Nc-1]\big]}_{\wbZ}, \label{equ:20} \\
	& = \bPhi \big((\olbCT\olbGammaT\olbAT(\btheta))\star \bCR\bGammaR\bAR(\bphi)\big)\underbrace{\big[\bg[0],\ldots,\bg[\Nc-1]\big]}_{\wbG} +\wbZ,\label{equ:crlb20}
	\end{align}
	where $\wbY\in\bbC^{M\Lr \times \Nc}$, 
	$\wbH\in\bbC^{ \Kr \Kt\times \Nc}$, $\wbZ\in\bbC^{M\Lr\times \Nc}$ and $\wbG\in\bbC^{\Np\Nray \times \Nc}$ are defined accordingly.
	Vectorizing \eqref{equ:crlb20}, we have
	\begin{align} 
	\vec(\wbY)  
	& =	\Big(\bI_{\Nc}\kron \bPhi\big(\olbCT\olbGammaT \olbAT(\btheta) \star \bCR\bGammaR\bAR(\bphi)\big)\Big)\vec(\wbG) + \vec(\wbZ), \notag\\
	& = \Big(\bI_{\Nc}\kron \bPhi \big(\olbCT\kron \bCR\big)\big(\olbGammaT\kron \bGammaR \big)\big(\olbAT(\btheta) \star \bAR(\bphi)\big)\Big)\vec(\wbG) + \vec(\wbZ).
	\end{align} 
	Therefore, the received signal is Gaussian distributed as $\vec(\wbY)\sim\cC\cN(\bmu,\bC_{\vec(\wbZ)})$, where $\bmu \teq \big(\bI_{\Nc}\kron \bPhi \big(\olbCT\kron \bCR\big)\big(\olbGammaT\kron \bGammaR \big)\big(\olbAT(\btheta) \star \bAR(\bphi)\big)\big)\vec(\wbG)$ and \sloppy $\bC_{\vec(\wbZ)} \teq  \sigma^2\bI_{\Nc}\kron \text{blkdiag}\big\{\bW_0^*\bW_0, \bW_1^*\bW_1,\cdots,\bW_{M-1}^*\bW_{M-1}\big\}$.
	For simplicity, we can further whiten the received signal by $\sigma\bC_{\vec(\wbZ)}^{-1/2} \vec(\wbY)$, such that the whitened mean is $\bmuw \teq \big(\bI_{\Nc}\kron \bPhiw \big(\olbCT\kron \bCR\big)\big(\olbGammaT\kron \bGammaR \big)\big(\olbAT(\btheta) \star \bAR(\bphi)\big)\big)\vec(\wbG)$, with $\bPhiw\teq  \text{blkdiag}\big\{\bW_0^*\bW_0, \bW_1^*\bW_1,\cdots,\bW_{M-1}^*\bW_{M-1}\big\}^{-1/2}\bPhi$, and the whitened noise covariance matrix is $\sigma^2\bI_{M\Lr\Nc}$.

Let us stack all the unknown parameters of interest into a vector $\bxi$, including $\Np\Nray$ pairs of AOAs/AODs $\{\bphi,\btheta\}$, the real and imaginary parts of the path gain vectors $\bg^{\R}[c], \bg^{\I}[c]$ (from $\bg[c]\teq\bg^{\R}[c]+\j \bg^{\I}[c], \forall c\in\cI(\Nc)$), the antenna gain errors $\bm g_{\t}\teq\{g_{\t,i}\}_{i=1}^\Nt$ and $\bm g_{\r}\teq\{g_{\r,i}\}_{i=1}^\Nr$, the antenna phase errors $\bm \phi_{\t}\teq\{\phi_{\t,i}\}_{i=1}^\Nt$ and $\bm \phi_{\r}\teq\{\phi_{\r,i}\}_{i=1}^\Nr$, the antenna spacing errors $\bm\epsilon_{\t}\teq\{\epsilon_{\t,i}\}_{i=1}^{\Nt-1}$ and $\bm\epsilon_{\r}\teq\{\epsilon_{\r,i}\}_{i=1}^{\Nr-1}$, and the coupling coefficients $\bm c_{\t}\teq \{ c_{\t,i,j},\forall  1{<}i{<}j{<}\Nt\}$ and $\bm c_{\r}\teq\{ c_{\r,i,j}, \forall 1{<}i{<}j{<}\Nr\}$. Note that there are only $\Nt-1$ and $\Nr-1$ spacing error terms at the transmitter and receiver, respectively. Moreover, there are  $\frac{\Nt(\Nt-1)}{2}$ and $\frac{\Nr(\Nr-1)}{2}$ unknown coupling coefficients in $\bCT$ and $\bCR$, respectively, for the coupling matrix for ULAs is Toeplitz matrices with diagonal elements being one (notice that the number of coupling coefficients is different if other types of coupling matrices are assumed). 
Therefore, 
\sloppy $\bxi\teq[\bm\theta^T,\bm\phi^T, \bg^{\R}[0]^T,\bg^{\I}[0]^T,\ldots,\bg^\R[\Nc{-}1]^T,\bg^\I[\Nc{-}1]^T,\bm g_{\t}^T,\bm g_{\r}^T,\bm\phi_{\t}^T,\bm\phi_{\r}^T,\bm\epsilon_{\t}^T,\bm\epsilon_{\r}^T,\bm c_{\t}^T,\bm c_{\r}^T]^T\in\bbC^{2\Np\Nray+2\Np\Nray\Nc+2\Nr+2\Nt+(\Nt-1)+(\Nr-1)+\Nr(\Nr-1)/2+\Nt(\Nt-1)/2}$.
Then the $(i,j)$-th entry in the overall Fisher Information Matrix (FIM) for $\bxi$, denoted by $[\bI(\bxi)]_{\xi_i,\xi_j}$, is given as \cite{Kay:Fundamentals-of-statistical-signal:93}
\begin{align} \label{equ:FIM} 
[\bI(\bxi)]_{\xi_i,\xi_j}=2\Re\left\{\frac{\partial \bmu^*(\bxi)}{\partial \xi_i}\bC_{\vec(\wbZ)}^{-1}(\bxi)\frac{\partial \bmu(\bxi)}{\partial \xi_j}\right\}
=\frac{2}{\sigma^2}\Re\left\{\frac{\partial \bmuw^*(\bxi)}{\partial \xi_i} \frac{\partial \bmuw(\bxi)}{\partial \xi_j}\right\}.
\end{align}  
Based on \eqref{equ:FIM}, the detailed derivations of FIM $\bI(\bxi)$ are given in Appendix.
Once the FIM $\bI(\bxi)$ in \eqref{equ:FIMfinal} is available, considering that the channel  $\vec(\bH[c])=\big((\olbCT\olbGammaT\olbAT(\btheta))\star \bCR\bGammaR\bAR(\bphi)\big)\bg[c]$ is a function of the unknown variable $\bxi$, then the covariance matrix of any unbiased estimator for  $\vec(\bH[c])$ is given as \cite{Kay:Fundamentals-of-statistical-signal:93}
\begin{align} \label{equ:Jacobian} 
\bC_{\vec(\bH[c])}\geq \bJ_{\bxi}(c)\bI^{-1}(\bxi)\bJ_{\bxi}^*(c), \quad \forall c\in\cI(\Nc),
\end{align}
where $\displaystyle\bJ_{\bxi}(c)\teq \frac{\partial\vec(\bH[c]) }{\partial \bxi^T}$ is the Jacobian matrix of the channel $\vec(\bH[c])$ w.r.t. $\bxi$. It can be expressed as a block column matrix as $	\bJ_{\bxi}(c) = [\bJ_{\btheta}(c) \ \bJ_{\bphi}(c)\ \bJ_{\bg[c]}(c) \ \bJ_{\bm g_{\t}}(c) \ \bJ_{\bm g_{\r}}(c) \ \bJ_{\bm \phi_{\t}}(c) \ \bJ_{\bm \phi_{\r}}(c)\ \bJ_{\bm \epsilon_{\t}}(c) \ \bJ_{\bm \epsilon_{\r}}(c) \ \bJ_{\bm c_{\t}}(c)\ \bJ_{\bm c_{\r}}(c)]$, and the expressions of each column block can be derived similarly following the above procedure of derivation for the FIM $\bI(\bxi)$. Finally, the total CRLB for the channel matrices $\vec(\bH[c]), \forall c \in\cI(\Nc)$ is computed as
\vspace*{-1mm}\begin{align}
	\text{CRLB}= \sum_{c=0}^{\Nc-1} \tr\{\bJ_{\bxi}(c)\bI^{-1}(\bxi)\bJ_{\bxi}^*(c)\}.
\end{align}
The value of the CRLB will be included in the simulations as a reference to evaluate the channel estimation error performance when using the learned dictionaries to model the channel.

\vspace*{-2mm}
\section{Combined Dictionary Learning (CoDL)}\label{sec:CoDL}

In this section, we develop a solution to the problem of learning a combined dictionary for frequency-selective mmWave MIMO channels, i.e., the Kronecker product of transmit and receive dictionaries in \eqref{equ:Hkvect}.

\subsection{Basic problem formulation} 

Inspecting \eqref{equ:Mframe}, there are two unknown parameters, i.e., the combined dictionary $\bPsi$ and the channel vector $\wbh[c]$. 
The conventional DL problem for recovering the optimal $\bPsi$ as well as $\wbh[c]$ can be formulated as  
\begin{align}\label{equ:DLprob3}
	\min_{\substack{\bPsi \in\cD,~ \wbh[c] }}\ \ \big\|\wby[c] - \bPhi\bPsi\wbh[c]\big\|_2^2 + w\big\|\wbh[c]\big\|_0,  
\end{align}
where  $\|\cdot\|_0$ is the $\ell_0$-norm of a vector, i.e., the number of non-zero elements and $w$ is a regularization factor that trades off the model fitting error and sparsity level.  
Moreover, the unit-norm condition in \eqref{equ:HdDL} transfers to the unit-norm constraint for the combined dictionary as 
\begin{align}
\cD = \Big\{\bPsi\in\bbC^{\Nt\Nr \times \Kr \Kt} \big| \|[\bPsi]_{:,j}\|_2= 1,\ \forall j\in\cI(\Kr \Kt)\Big\}. 
\end{align}  

\subsection{Advanced problem formulation}

Beyond the basic formulation in  \eqref{equ:DLprob3}, the DL problem can be extended by considering the  following aspects:

\subsubsection{Common support among subcarriers}
The frequency domain channel vectors $\wbh[c], \ c\in\cI(\Nc)$ exhibit an important property: common sparsity support among subcarriers  \cite{RodGonVen:Frequency-domain-Compressive-Channel:18}. Specifically, the rows of $\wbH$ in \eqref{equ:20} will be either all zeros or non-zeros.
This property is commonly used when beam squint effect is not considered. 

\subsubsection{Multiple measurements at different locations}

In addition to increasing the number of training OFDM symbols and exploiting the common support between subcarriers, it is possible for the UE to collect multiple measurements at different positions. 
This way, a larger training set is created where different channel sparsity patterns are included, and then the sparsifying dictionary does not depend on the specific location, but on the specific impairments. Note that this procedure can be done at the stage of network setup, for instance, in an indoor WiFi scenario.  
Suppose there are $\Nsa$ measurement samples collected at different positions, and each measurement $\wbY$ defined in \eqref{equ:20} is then denoted by $\wbY^{(u)},u\in\cI(\Nsa)$.  Stacking them in a compact form gives   
\begin{align}\label{equ:23}  
	\underbrace{\big[\wbY^{(0)}, \wbY^{(1)},\ldots,\wbY^{(\Nsa-1)}\big]}_{\bsfY} 
	&= \bPhi\bPsi
	\underbrace{\big[\wbH^{(0)},\wbH^{(1)},\ldots,\wbH^{(\Nsa-1)}\big]}_{\bsfH} + \underbrace{\big[\wbZ^{(0)},\wbZ^{(1)},\ldots,\wbZ^{(\Nsa-1)}\big]}_{\bsfZ},
\end{align}
where $\bsfY\in\bbC^{M\Lr\times \Nc \Nsa}$, $\bsfH\in\bbC^{\Kr \Kt\times \Nc \Nsa}$ and
$\bsfZ\in\bbC^{M\Lr\times \Nc \Nsa}$ are defined accordingly. 

\subsubsection{Denoising option} The proposed DL algorithm has to be able to operate under the low SNR conditions in mmWave communications. Next, we denote a denoised (and unknown) version of $\bsfY$ in \eqref{equ:23} by $\bsfX$, i.e., $\bsfY=\bPhi \bPsi\bsfH + \bm{\mathsf Z} = \bsfX+\bm{\mathsf Z'} $, where $\bsfZ'$ is comprised of $\bsfZ$ and the mismatch error between $\bsfX$ and $\bPhi\bPsi\bsfH$. Then, an additional regularization term $\|\bsfY-\bsfX\|_F^2$ can be added to the following DL problem formulation to alleviate the influence of noise. 

Given these three aspects, the CoDL problem in \eqref{equ:DLprob3} can be written as \cite{ElaAha:Image-Denoising-Via-Sparse:06}
\begin{align}\label{equ:commmulti}
\min_{\substack{\bPsi\in\cD, \  \bsfX, \\ \wbhu[0],\ldots, \wbhu[\Nc-1]}}\ \ &\big\|\bsfX- \bPhi\bPsi\bsfH\big\|_F^2+w_1\sum_{u=0}^{\Nsa-1}\sum_{c=0}^{\Nc-1}\big\|\wbhu[c]\big\|_0 +w_2\|\bsfY-\bsfX\|_F^2 \notag\\
\text{s.t.}\quad \quad \ \quad 
& \supp\{\wbhu[0]\} =\cdots 
=\supp\{\wbhu[\Nc-1]\},\ \forall u\in\cI(\Nsa).
\end{align}
Note that an additional superscript $(u)$ is added as sample index for $\wbhu[c]$ to indicate which channel matrix $\wbHu$ it belongs to. 
Inspecting \eqref{equ:commmulti}, the sparsity enhancement regularization term and the common sparsity support property can be considered jointly, i.e., using the joint sparsity regularization $\ell_2/\ell_1$ norm of $\wbHu, \forall u\in\cI(\Nsa)$, rather than encoding each subcarrier channel vector $\wbhu[c]$ separately. In doing so, the sparsity enhancement regularization term together with the common sparsity support constraint in \eqref{equ:commmulti} can be integrated as $f(\bsfH)$, given by
\begin{align}
f(\bsfH) = \sum_{u=0}^{\Nsa-1} \big\|\wbHu\big\|_{2,1}= \sum_{u=0}^{\Nsa-1} \sum_{i=0}^{\Kr \Kt-1} \big\|\big[\wbHu\big]_{i,:}\big\|_2. 
\end{align} 
Then, we have the final formulation for the CoDL problem as  
\begin{align}\label{equ:SDLL}
	\min_{\bPsi\in\cD,\ \bsfH,\ \bsfX} \  \big\|\bsfX - \bPhi\bPsi\bsfH\big\|_F^2 + w_1 f(\bsfH)+w_2\|\bsfY-\bsfX\|_F^2.
\end{align} 

\subsection{Optimization for the CoDL  problem}\label{sec:CoDL_optimization}

The objective function in \eqref{equ:SDLL} is not jointly convex w.r.t. $ \bPsi$, $\bsfH$ and $\bsfX$, but it is convex in each of the variables when the others are kept fixed. 
Thereby, a possible approach to finding the solution of \eqref{equ:SDLL} involves solving three sub-problems:
1) updating the channel $\bsfH$ with fixed $\bPsi$ and $\bsfX$, 2) updating the dictionary $\bPsi$ with fixed $\bsfX$ and newly updated $\bsfH$, and 3) updating $\bsfX$ with fixed newly updated $\bsfH$ and $ \bPsi$. 

\subsubsection{Sparse coding stage} 

Let us first assume that $\bPsi$ and $\bsfX$ are fixed, so that the optimization problem is reduced to a sparse coding problem for updating $\bsfH$. Thanks to the separability of the objective function \eqref{equ:SDLL} w.r.t. each $\wbHu, u\in\cI(\Nsa)$, 
we can compute $\wbHu$ separately. Specifically, all
$\wbH^{(v)},~ v\neq u$ are fixed when computing $\wbHu$. Along these lines, the
objective function in \eqref{equ:SDLL} can be further simplified to
\begin{align}\label{equ:28}
	\min_{\wbHu} \ \big\|\wbX^{(u)}-\bPhi \bPsi\wbHu\big\|_F^2 + w_1\big\|\wbHu\big\|_{2,1}. 
\end{align}  
Various sparse coding techniques can be used to solve this problem, such as the orthogonal matching pursuit (OMP) \cite{TroGil:Signal-Recovery-From:07}, the simultaneous-weighted (SW-OMP) \cite{RodGonVen:Frequency-domain-Compressive-Channel:18},
and the alternating direction method of multipliers (ADMM) \cite{LiuCheChe:Support-Discrimination-Dictionary:16,BoyParChu:Distributed-Optimization-and-Statistical:11,YanZha:Alternating-Direction-Algorithms:11}. 
\subsubsection{Dictionary update stage}

Following the sparse coding step, we then update dictionary $ \bPsi$ column by column with fixed $\bsfX$ and newly estimated $\bsfH$. When updating the $k$-th ($k\in\cI(\Kr \Kt)$) atom, all the other columns are fixed as well. Then the dictionary update is formulated as 
\begin{align}\label{equ:CoDLdictionaryupdate}
	\min_{\bPsi }\ \big\|\bsfX-\bPhi\bPsi\bsfH\big\|_F^2,\quad \text{s.t.}\ \  \big\|[ \bPsi]_{:,k}\big\|_2=1.
\end{align} 
This problem can be solved by using the well-known K-SVD algorithm \cite{AhaElaBru:K-svd-An-algorithm-for-designing:06} (or approximate K-SVD \cite{RubZibEla:Efficient-implementation-of-the-K-SVD:08} with reduced complexity), which updates the dictionary $\bPsi$ atom by atom, or by the canonical method of optimal direction (MOD) algorithm \cite{EngAasHus:Method-of-optimal-directions:99}. 

\begin{algorithm}[t]
	\caption{: Combined dictionary learning (CoDL)}
	\label{alg::ADMMKSVD}
	\begin{itemize}
		\item \textbf{Input:} Training measurements $\bsfY\in\bbC^{M\Lr\times \Nc \Nsa}$, measurement matrix $\bPhi\in\bbC^{M\Lr\times \Nr \Nt}$, number of OFDM subcarriers $\Nc$, number of measurement samples $\Nsa$, regularization parameters $w_1,w_2$. 
		\item \textbf{Initialization:} Set the dictionary matrix $\bPsi\in\bbC^{\Nt\Nr\times \Kr \Kt}$ using measurement data with normalized columns, and set $\bsfX=\bsfY$.
		\item \textbf{While} \textit{not converge} \textbf{do}
		\item[] \quad \quad 1. \textit{Sparse coding stage}: Solve \eqref{equ:28} for each $u\in\cI(\Nsa)$ to get the channel matrix $\bsfH=\big[\wbH^{(0)},\wbH^{(1)},\ldots,\wbH^{(\Nsa-1)}\big]$.
		\item[] \quad \quad 2. \textit{Dictionary update stage}: If MOD is adopted, the update of $\bPsi$ is given by 
		$
			\bPsi = \bPhi^{\dag}\bsfX\bsfH^*(\bsfH\bsfH^*)^{-1}
		$, followed by normalization on each column.  
		If K-SVD is considered, $\bPsi$ is updated column by column as did in \cite{AhaElaBru:K-svd-An-algorithm-for-designing:06,RubZibEla:Efficient-implementation-of-the-K-SVD:08}.
		\item[] \quad \quad 3. \textit{Denoising stage}: Update $\bsfX$ by solving \eqref{equ:Xupdate}, i.e., $[\bsfX]^{\new}=(1+w_2)^{-1}(w_2\bsfY+\bPhi \bPsi\bsfH)$.
		\item[] \textbf{end while} 
		\item \textbf{Output:} The optimal dictionary $\bPsi$, the channel matrix $ {\bsfH}$ and the denoised version measurement $ {\bsfX}$.
	\end{itemize} 
\end{algorithm} 

\subsubsection{Denoising stage}

Following the sparse coding and dictionary update stages, the value of $\bsfX$ is updated by computing
\begin{align}\label{equ:Xupdate}
\min_{ \bsfX} \  \big\|\bsfX - \bPhi\bPsi\bsfH\big\|_F^2 +  w_2\|\bsfY-\bsfX\|_F^2,
\end{align} 
which can be solved by least squares (LS) and given as $[\bsfX]^{\new}=(1+w_2)^{-1}(w_2\bsfY+\bPhi \bPsi\bsfH)$.

The whole procedure of the proposed CoDL algorithm is summarized in \algoref{alg::ADMMKSVD}. 
The proposed algorithm will stop either if the values of objective function  \eqref{equ:SDLL} at adjacent iterations are sufficiently close, or if the maximum number of iterations is reached. In \sref{sec:Simulations}, we will show via numerical simulations that the objective function is to decrease quickly as the number of iterations increase, which helps to illustrate fast convergence of the proposed algorithms. 

\subsection{Convergence analysis and dictionary initialization}

For sparse coding stage, the ADMM algorithm can compute the exact solution for each sub-problem, its convergence is guaranteed by the existing ADM theory \cite{GloBad:Numerical-methods-for-nonlinear:85,LiuCheChe:Support-Discrimination-Dictionary:16}. In this stage, when the dictionaries are fixed, each sparse coding step decreases the value of the objective function.  
While for the dictionary update and denoising stages, as explained in  \cite{AhaElaBru:K-svd-An-algorithm-for-designing:06}, an additional reduction or no change in the mismatch error is guaranteed. Therefore, the alternating steps for optimizing the CoDL problem ensure a monotonic decrease in the objective function and then convergence to a local minimum is guaranteed.

As the alternating optimization can only guarantee to converge to a local minimum, the initialization is then of significant importance to avoid local minimizers and ensure the learned dictionaries to be closer to the actually dictionaries. The common initialization choices for current DL algorithms include random initial dictionary, an overcomplete wavelet/Fourier dictionary or a sample of data measurements. Some other initialization methods were also proposed for different DL problems in the literature, e.g., \cite{RusDum:An-initialization-strategy-for-the-dictionary:14,AgaAnaNet:A-clustering-approach-to-learning:17,ChaBar:Alternating-minimization-for-dictionary:17}.  In this paper, after thorough comparisons and investigations among different initialization methods, we finally adopt the dictionary initialization algorithm (DIA) proposed in  \cite{RusDum:An-initialization-strategy-for-the-dictionary:14}. The main idea of DIA is to use incoherent structures to create a very good initialization for a DL problem, which involves an iterative adaptation of the dictionary to the dataset with pruning of the less used atoms and constructions of new atoms that fit the data better. The detailed procedures of DIA can be found in the Algorithm 1 in \cite{RusDum:An-initialization-strategy-for-the-dictionary:14}.

\section{Separable Dictionary Learning (SeDL)}\label{sec:SeDL}

Inspecting \eqref{equ:SDLL}, the combined dictionary $\bPsi$ was learned without considering the specific structure constraints on $ \bDT$ and $ \bDR$, i.e., $\|\bd_{\T,k_t}\|_2=1$, $\forall k_t\in\cI(\Kt)$ and  $\|\bd_{\R,k_r}\|_2=1$, $\forall k_r\in\cI(\Kr)$. In this section, we investigate the formulation and optimization for the SeDL problem. By separating the constraints on transmit and receive dictionaries, the SeDL problem will be more suited for the practical MIMO systems. 

\subsection{Problem formulation}

\begin{figure}[t]
	\centering
	\includegraphics[width=120mm]{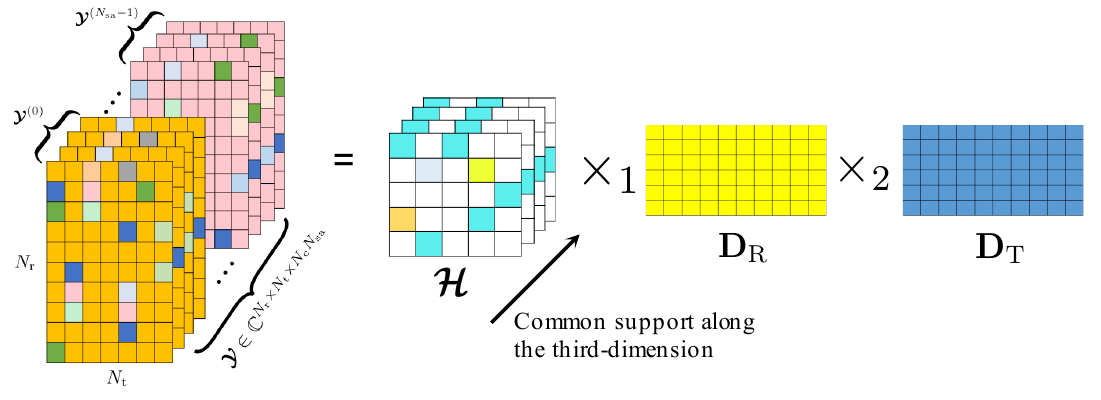}
	\caption{Illustration of tensor expression of $\bcY\in\bbC^{\Nr\times \Nt\times \Nc \Nsa}$ and its tensor mode product expansion.}\label{fig:tensorillustration} 
\end{figure}

To facilitate the problem formulation for SeDL, we re-formulate \eqref{equ:23} in a higher dimensional tensor space \cite{KolBad:Tensor-Decompositions-and-Applications:09}. Specifically, the collected measurements after removing the training sequences, i.e., $\bPhi^{\dag}\bsfY\in\bbC^{\Nr \Nt\times \Nc \Nsa}$ in \eqref{equ:23}, is re-written as a three-order (three-dimensional) tensor  $\bcY\in\bbC^{\Nr\times \Nt\times \Nc \Nsa}$, 
which is stacked by $\Nsa$ sub-tensors  $\bcY^{(u)}\in\bbC^{\Nr\times \Nt\times \Nc}, u\in\cI(\Nsa)$ along the third dimension, as shown in \figref{fig:tensorillustration}. A similar definition goes for its unknown denoised version $\bcX\in\bbC^{\Nr\times \Nt\times \Nc \Nsa}$. We also replace $\bsfH\in\bbC^{\Kr \Kt\times \Nc \Nsa}$ with a tensor  $\bcH\in\bbC^{\Kr\times \Kt\times  \Nc \Nsa}$, 
which is stacked by $\Nsa$ sub-tensors  $\bcHu\in\bbC^{\Kr\times \Kt\times \Nc}, u\in\cI(\Nsa)$ along the third dimension. Then we have the following formulation for the SeDL problem as  
\begin{align}\label{equ:SeDL} 
\min_{\bcH,\bDR,\bDT} \quad 
& \sum_{u=0}^{\Nsa-1}\big\|\bcXu-{\bcHu}_{\times 1~}{ \bDR}_{~\times  2~}\olbDT\big\|_F^2 + w_1  \sum_{u=0}^{\Nsa-1}\|\bcHu\|_{1,1,2} +
	w_2\big\|\bcY-\bcX\big\|_F^2\notag\\
\text{s.t.}\quad \ \ &\|\bd_{\T,k_t}\|_2=1,\ \|\bd_{\R,k_r}\|_2=1,\ \forall k_t\in\cI(\Kt), k_r\in\cI(\Kr).   
\end{align} 
where $\|\bcHu\|_{1,1,2}$ denotes the summation of $\ell_2$ norm of all the mode-3 columns along the third dimension in $\bcHu$ and is used 
to promote the common sparsity support between subcarriers.

\subsection{Optimization for the SeDL problem}

Similar to \sref{sec:CoDL_optimization}, the optimization for SeDL in \eqref{equ:SeDL} is also divided into three parts, i.e., 1) sparse coding, 2) dictionary update, and 3) denosing stage.

\subsubsection{Sparse coding stage}
Let us first assume that the transmit/receive dictionaries, $ \bDT$ and $ \bDR$, together with $\bcX$
are fixed, and then update the channel tensor $\bcHu$ for each $u\in\cI(\Nsa)$. Under these assumptions, the objective function is reduced to  
\begin{align}\label{equ:tensor1} 
&\min_{\bcHu}\quad
\big\|\bcXu-{\bcHu}_{\times 1~}{ \bDR}_{~\times 2~}\olbDT\big\|_F^2 + 
w_1 \big\|\bcHu\big\|_{1,1,2}. 
\end{align} 
After some mathematical manipulations, \eqref{equ:tensor1} is equivalent to
\begin{align}\label{equ:opt1}
&\min_{\bcHu}\quad  \left\|{[\bcXu]}_{(3)}^T- \left(\olbDT\kron \bDR\right){[\bcHu]}_{(3)}^{T}\right\|_F^2 + w_1 \left\|{[\bcHu]}_{(3)}^{T}\right\|_{2,1},
\end{align} 
where the $\ell_{1,1,2}$ norm of tensor is simplified to the $\ell_2/\ell_1$ norm of matrix according to the definitions of these norms and the involved tensor operations. Inspecting  \eqref{equ:opt1}, it is exactly in the form of \eqref{equ:28}. In fact, this is intuitive that when $ \bDT$ and $ \bDR$ are fixed, they can be combined to replace $ \bPsi$ in \eqref{equ:28} and thereby the update of $\bcHu$ is similar to that for $\wbH^{(u)}$. Note that besides transforming back to the form of \eqref{equ:28}, the sparse coding algorithms, e.g.,  ADMM, can be re-derived for this SeDL problem, which leads to lower computational complexity for the smaller dimensions of $\olbDT$ and $\bDR$ compared to the Kronecker product $\left(\olbDT\kron \bDR\right)$.

\subsubsection{Dictionary update stage} 

Following the sparse coding stage, the two dictionaries $ \bDT$ and $ \bDR$ are then updated column by column with fixed  $\bcH$ and $\bcX$. Note that rather than considering a combined dictionary $ \bPsi$ as did in \sref{sec:CoDL}, the update of $ \bDT$ and $ \bDR$ here are separated. 
If MOD is considered, we first fix $ \bDT$ and then the optimization of \eqref{equ:SeDL} w.r.t. $ \bDR$ is equivalent to 
\begin{align} \label{equ:SeDLMOD}
\min_{ \bDR}\quad \left\|{[\bcX]}_{(1)}- \bDR{[\bcH]}_{(1)}(\bI_{\Nsa}\kron \bI_{\Nc}\kron \olbDT)^T\right\|_F^2,
\end{align}  
where  ${[\bcX]}_{(1)}\teq\big[{[\bcX^{(0)}}]_{(1)},\cdots,{[\bcX^{(\Nsa-1)}]}_{(1)}\big]\in\bbC^{\Nr\times \Nt \Nc \Nsa}$ and its columns are the mode-1 fibers of $\bcX$. Similarly, ${[\bcH]}_{(1)}\teq\big[{[\bcH^{(0)}]}_{[(1)},\cdots,{[\bcH^{(\Nsa-1)}]}_{(1)}\big]\in\bbC^{\Kr\times \Kt\Nc \Nsa}$ collects the mode-1 fibers of $\bcH$. 
Therefore, the updates for $ \bDR$ is readily available by LS as
$ \label{equ:SeDLMOD1}
\big[ \bDR\big]^{\new} = [\bcX]_{(1)}\Big({[\bcH]}_{(1)}\big(\bI_{\Nsa}\kron \bI_{\Nc}\kron \big[\olbDT\big]^{\old}\big)^T\Big)^{\dagger}.
$
Similarly, the update for $\olbDT$ is given as  
$\label{equ:SeDLMOD2}
\big[\olbDT\big]^{\new} = {[\bcX]}_{(2)}\Big({[\bcH]}_{(2)}\big(\bI_{\Nsa}\kron \bI_{\Nc}\kron \big[\bDR\big]^{\new}\big)^T\Big)^{\dagger}. 
$
Besides, the basic idea of K-SVD for dictionary update could be extended accordingly in the tensor case by using the high-order SVD (HOSVD) \cite{RoeGalHaa:Tensor-based-algorithms-for-learning:14}.

\subsubsection{Denoising stage}

Following the sparse coding and dictionary update stages, to update $\bcXu,\forall u\in\cI(\Nsa)$, we can still transform it back to the denoising stage of CoDL, i.e.,
\begin{align} 
&\min_{\bcXu}\quad  \left\|{[\bcXu]}_{(3)}^T- \left(\olbDT\kron \bDR\right){[\bcHu]}_{(3)}^{T}\right\|_F^2 + w_2 \left\|[{\bcY^{(u)}]}_{(3)}^{T}-{[\bcXu]}_{(3)}^{T}\right\|_F^2,
\end{align} 
whose solution is also readily available by LS.

Based on these sparse coding, dictionary update and denoising steps, the whole procedure for SeDL can be summarized similar to \algoref{alg::ADMMKSVD}, which we omit for space limitation.    

\subsection{Complexity analysis}
  
The convergence analysis and stopping rules of the SeDL problem are similar to those of CoDL in \sref{sec:CoDL}. The analysis of computational complexity in terms of complex multiplication operations per iteration for CoDL and SeDL is provided in \tabref{tab:complexity}, where we assume $\Nr=\Nt=\sqrt{N}$ and $\Kr=\Kt=\sqrt{K}$  for ease of comparison, and $S_0$ denotes the sparsity level of coefficient matrices.
Note that the approximate K-SVD \cite{RubZibEla:Efficient-implementation-of-the-K-SVD:08} can be used to alleviate the computational burden of K-SVD based algorithms. Moreover, the involved computations for dictionary update are implemented offline so that they will not affect the complexity of sparse coding once the dictionary is learned.
\begin{table}[t]
	\centering
	\caption{Complexity analysis for different sparse coding and dictionary update algorithms}\label{tab:complexity}
	\begin{tabular}{|c|cc|c|c|} 
		\hline 
		&  \multicolumn{2}{c|}{\multirow{2}{*}{\parbox{3cm}{\centering Sparse coding \quad (online computation)}} } & \multicolumn{2}{c|}{Dictionary update (offline computation) } \\
		\cline{4-5} 
		& & & MOD & K-SVD/K-HOSVD \\ 
		\hline
		SW-OMP \cite{RodGonVen:Frequency-domain-Compressive-Channel:18}  & \multicolumn{2}{c|}{$\cO(\Nsa\Nc N K)$ } & \multicolumn{2}{c|}{ \quad \ \ ---}\\
		
		\hline
		\multicolumn{1}{|c|}{\multirow{2}{*}{CoDL}}   & \multicolumn{2}{c|}{\multirow{2}{*}{\parbox{3 cm}{\centering $\cO\big(\Nsa (NK^2 + K^3) +\Nsa\Nc(NK   + K^2)  \big) $}}} & \multirow{2}{*}{\parbox{3.5cm}{\centering $ \cO\big( \Nsa\Nc NK + NK^2  + \Nsa\Nc K^2 + K^3  \big)$  }}&  \multirow{2}{*}{\parbox{3.5cm}{\centering $\cO\big(N^2 (S_0\Nsa\Nc+NK  ) \big)  $ }}\\
		& & & & \\ 
		\hline
		\multicolumn{1}{|c|}{\multirow{2}{*}{SeDL}}  &\multicolumn{1}{c|}{Reducing to CoDL} &  The same as CoDL & \multirow{2}{*}{\parbox{4cm}{\centering $\cO\big(2\Nsa\Nc(N \sqrt{K}+K\sqrt{N})   + 6\Nsa\Nc K\sqrt{K} + 2K\sqrt{K} \big)   $}} & \multirow{2}{*}{\parbox{3cm}{\centering $\cO\big(N^2 (S_0\Nsa\Nc +NK )+ 2\Nsa\Nc N^3S_0   \big)$} }\\ 
		\cline{2-3} 
		&\multicolumn{1}{c|}{ADMM for SeDL} & $\cO\big(4\Nsa\Nc K\sqrt{K}\big)$ & &   \\
		\hline	 
	\end{tabular}  
\end{table}

\subsection{Channel estimation with learned dictionary}
Once the optimal combined dictionary $\bPsi$ or separable dictionaries $ \bDR$ and $ \bDT$ are learned via CoDL and SeDL algorithms, they will be used for channel estimation in the following transmissions. Then, the problem of channel estimation at UE exactly boils down to a compressive sensing (sparse coding) problem, as formulated in \eqref{equ:28} and \eqref{equ:tensor1} with fixed $\bPsi$ or $ \bDR$ and $ \bDT$, so that the various sparse coding algorithms for CoDL and SeDL can be directly applied, such as OMP, SW-OMP, ADMM, to name a few.

\section{Numerical Results}\label{sec:Simulations}

In this section, we provide numerical simulations to corroborate the effectiveness of the proposed DL and channel estimation algorithms for hybrid wideband mmWave  MIMO systems. For comparison, compressive channel estimation (sparse coding) by SW-OMP \cite{RodGonVen:Frequency-domain-Compressive-Channel:18} and the canonical OMP methods \cite{TroGil:Signal-Recovery-From:07}, using the overcomplete ideal array response matrix (IARM) uniformly sampled in the physical angle domain as the sparsifying dictionaries are also evaluated. 

\subsection{Simulation parameters} 
Unless otherwise specified, the default parameters in the simulations are summarized as follows. Both the AP and the UE deploy a ULA with (presumed) half-wavelength antenna spacing, and $\Nt=32$, $\Nr=8$, $\Ns=\Lt=\Lr=2$ and $\Kt=64$, $\Kr=16$. The phase-shifters used at the AP and the UE are assumed to have $\NQ=2$ quantization bits, so that the phases of the entries of precoders $\bF_m$ and combiners $\bW_m$ are randomly chosen from $\big\{0, \frac{2\pi}{2^{\NQ}},\ldots,\frac{2\pi(2^{\NQ}-1)}{2^{\NQ}}\big\}$. The number of OFDM subcarriers is set as $\Nc=128$ and $\Ts=\frac{1}{1760}~\mu s$, as specified in the IEEE 802.11ad wireless standard. 
The channels are generated according to \eqref{equ:Hd} with  $\Ntap=16$ delay taps and $\Np=6$ multipath components each of $\Nray=1$ ray (which is typical in indoor scenarios), whose delays are chosen uniformly from $[0,(\Ntap-1)\Ts]$. The band-limited filter $\prc(t)$ is assumed to be a raised-cosine filter with roll-off factor of $0.8$. 
Moreover, all the AOAs and AODs are supposed to be constrained in a sector range of $120^\circ$, because we expect to have multiple antenna array panels covering different sectors. The number of  measurement samples $\Nsa$ is set as $100$.  
The SNR is set to 0 dB for DL and channel estimation. 
During the DL phase, $M=500$ and the spreading factor 
$\Nrepeat = 10$ is used to increase the effective SNR in 10 dBs.

To evaluate the effects of array uncertainties on DL performances, we adopt the  models and parameters in  \cite{EbeEscBie:Investigations-on-antenna-array:16} to characterize the antenna gain and phase errors as well as their mutual coupling. More concretely, 
the gains and phase shifts in $\bGammaR$ and $\bGammaT$ are modeled as 
$g_i = 1.0 + 0.05\cdot\sigma_g$ and $\nu_i = \frac{20^\circ \pi}{180^\circ}\cdot \sigma_{\nu}$, with $\sigma_g\sim \cN(0,1)$ and $\sigma_{\nu}\sim \cN(0,1)$ being normal Gaussian distributions. This indicates that the gain and phase error variances for each antenna element are $5\%$ and $20^\circ\pi/180^\circ$. The mutual coupling coefficients among antennas are in the range between $0.01$ and $0.4$ as assumed in \cite{EbeEscBie:Investigations-on-antenna-array:16}. Moreover, to characterize the manufacture error and evaluate irregular array geometries, we introduce the antenna spacing perturbation as in \cite{DinRao:Dictionary-Learning-Based-Sparse:18}, and assume that the antenna spacing is not perfect $0.5\lambda$, but uniformly distributed between $0.4\lambda$ and $0.6\lambda$, with $\lambda$ being the carrier wavelength. 

For the optimization of CoDL and SeDL, unless otherwise specified, the ADMM-based algorithms are used for sparse coding, while approximate K-SVD based algorithms \cite{RubZibEla:Efficient-implementation-of-the-K-SVD:08} are used for dictionary update. The aforementioned DIA algorithm is used for dictionary initialization. 
The regularization parameters $w_1,w_2$ can be tuned and their default values are set as $w_1 = 0.1, w_2=0.001$. 
The learned dictionary is then used for subsequent channel estimation (i.e., solving the problem \eqref{equ:SDLL} and \eqref{equ:SeDL} with fixed learned dictionaries) by the ADMM-based sparse coding algorithm as well as the SW-OMP algorithm, while channel estimation by OMP and SW-OMP using the overcomplete IARM dictionary (with the same number of atoms as learned dictionaries) are also included. 

\subsection{Performances of DL and channel estimation with learned dictionaries}
\vspace*{-5mm}
\begin{figure}[!ht]  
	\centering	\includegraphics[width=80mm]{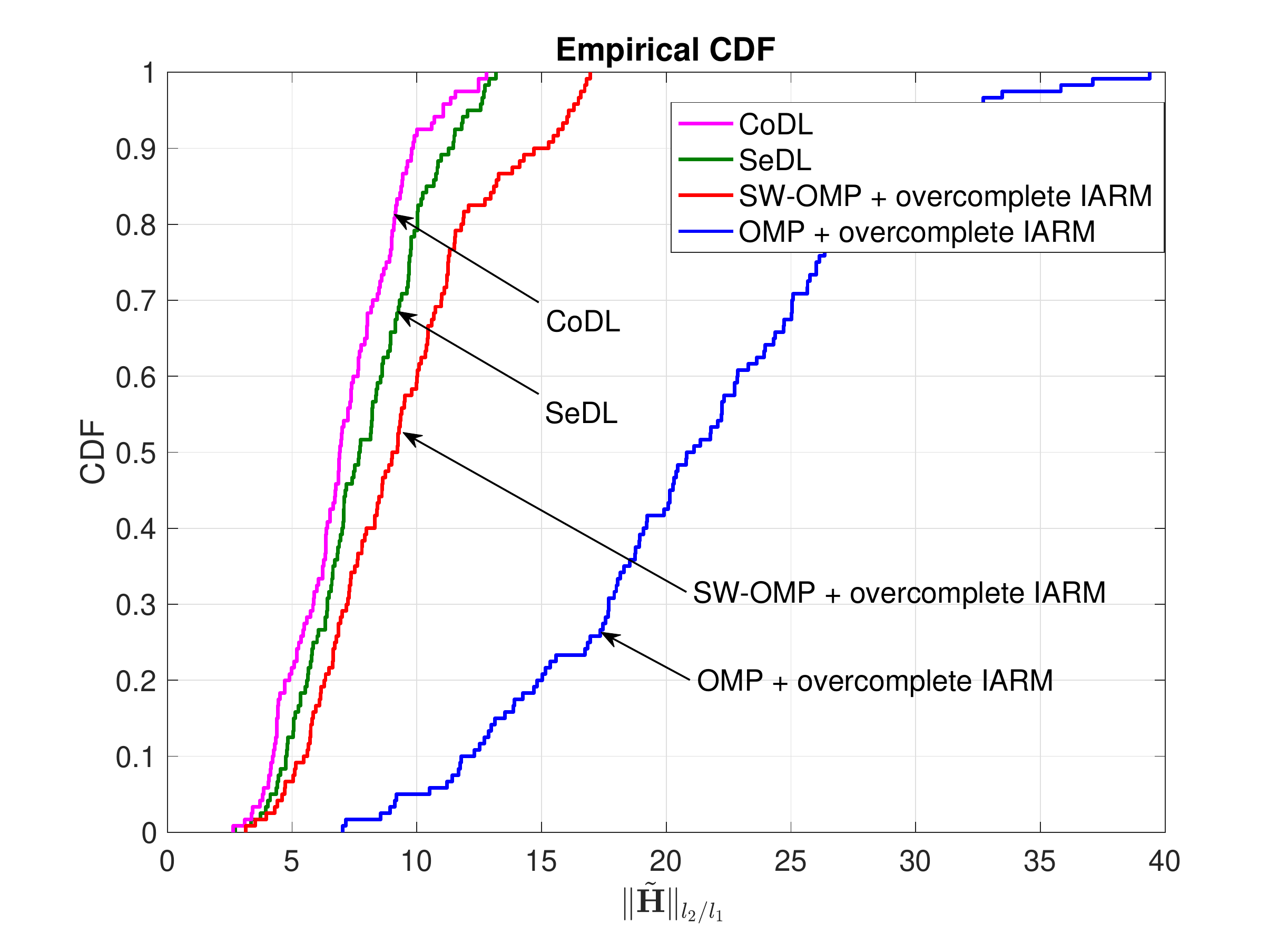}
	\vspace*{-5mm}
	\caption{The cumulative density function of $\|\wbH\|_{l_2/l_1}$ on different sparsifying dictionaries.}\label{fig:L2L1_MSE_M}  
\end{figure} 
\vspace*{-3mm}In \figref{fig:L2L1_MSE_M}, we compare the cumulative density functions (CDF) of $l_2/l_1$ norm of the estimated channel matrices, i.e., $\|\wbH\|_{l_2/l_1}$, using different sparsifying dictionaries, including learned dictionaries from CoDL and SeDL, as well as the overcomplete IARM dictionaries. As explained before, since the $l_2/l_1$ norm of a matrix is a convex surrogate of its row sparsity, CDFs of $\|\wbH\|_{l_2/l_1}$ on distinct dictionaries are then able to illustrate the capability of the corresponding dictionary for channel sparsity enhancement. It is clear from \figref{fig:L2L1_MSE_M}, the values of $\|\wbH\|_{l_2/l_1}$ corresponding to CoDL and SeDL are more likely smaller than those of overcomplete IARM dictionary, which means that the learned dictionaries by CoDL and SeDL can enhance the channel common sparsity compared to the overcomplete IARM dictionary. This corroborates the effectiveness of the learned dictionary for sparser channel representation in wideband systems. Moreover, comparing the two CDFs corresponding to overcomplete IARM dictionary using OMP and SW-OMP for sparse coding separately, it can be seen that the row sparsity of channel matrices is enhanced by SW-OMP. This is because the canonical OMP algorithm does not exploit the common sparsity properties of the wideband channel, and this suggests why the common sparsity support constraint is important and necessary in our proposed CoDL and SeDL algorithms. Lastly, the performance gap between CoDL and SeDL lies in the fact that the unit-norm constraint on the combined dictionary in \eqref{equ:SDLL} is generally less stricter than that on the separable transmit and receive dictionaries in \eqref{equ:SeDL}. In other words, the predetermined structure constraints on separable dictionaries limit their feasible ranges.

\begin{figure}[!ht] 
	\centering 
	\subfigure[Comparison of NMSE performances for ULA.]{
		\includegraphics[width=0.46\textwidth]{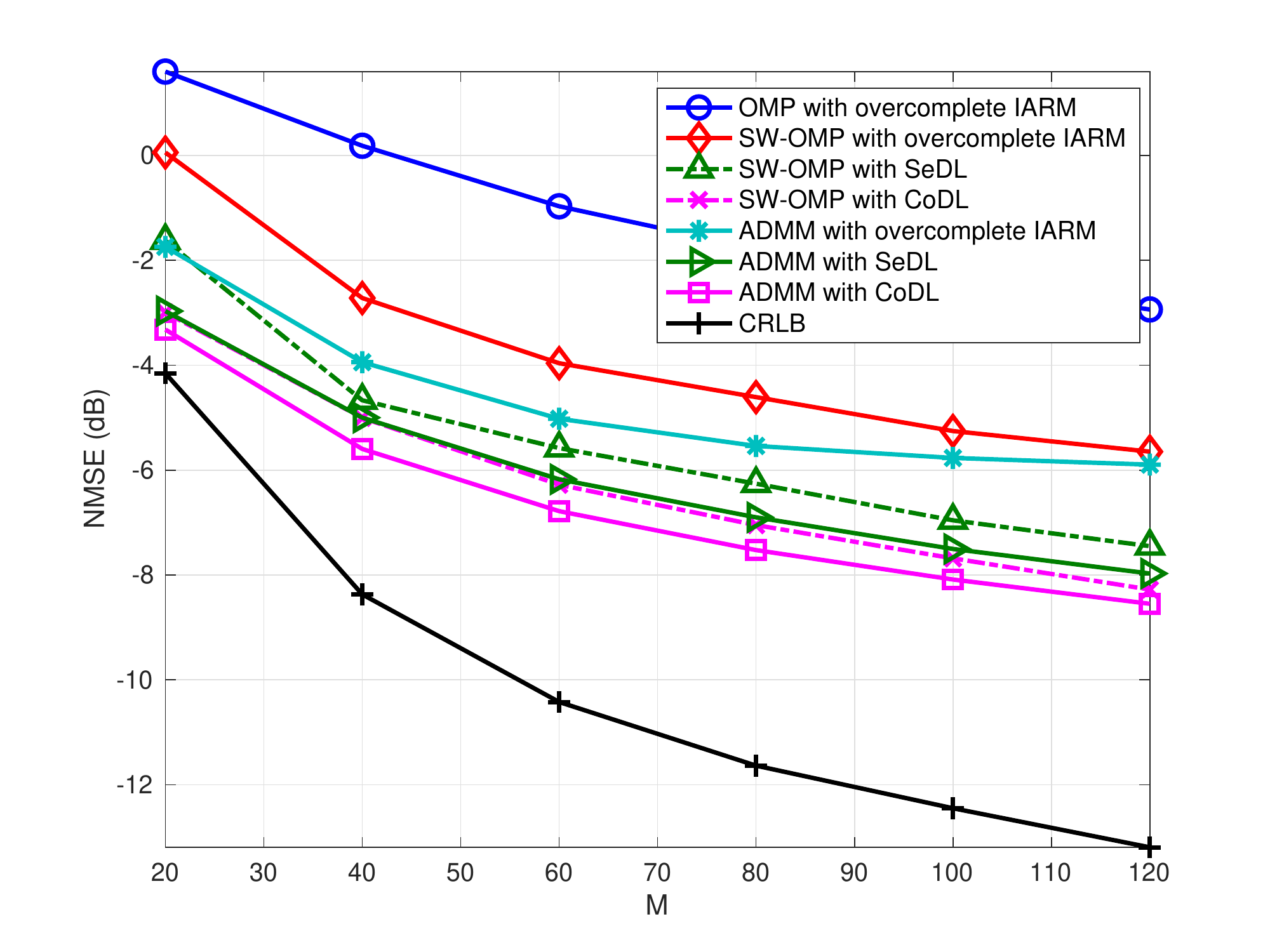}\label{fig:MSE_M}}
	\quad 
	\subfigure[Comparison of SE performances for ULA.]{
		\includegraphics[width=0.46\textwidth]{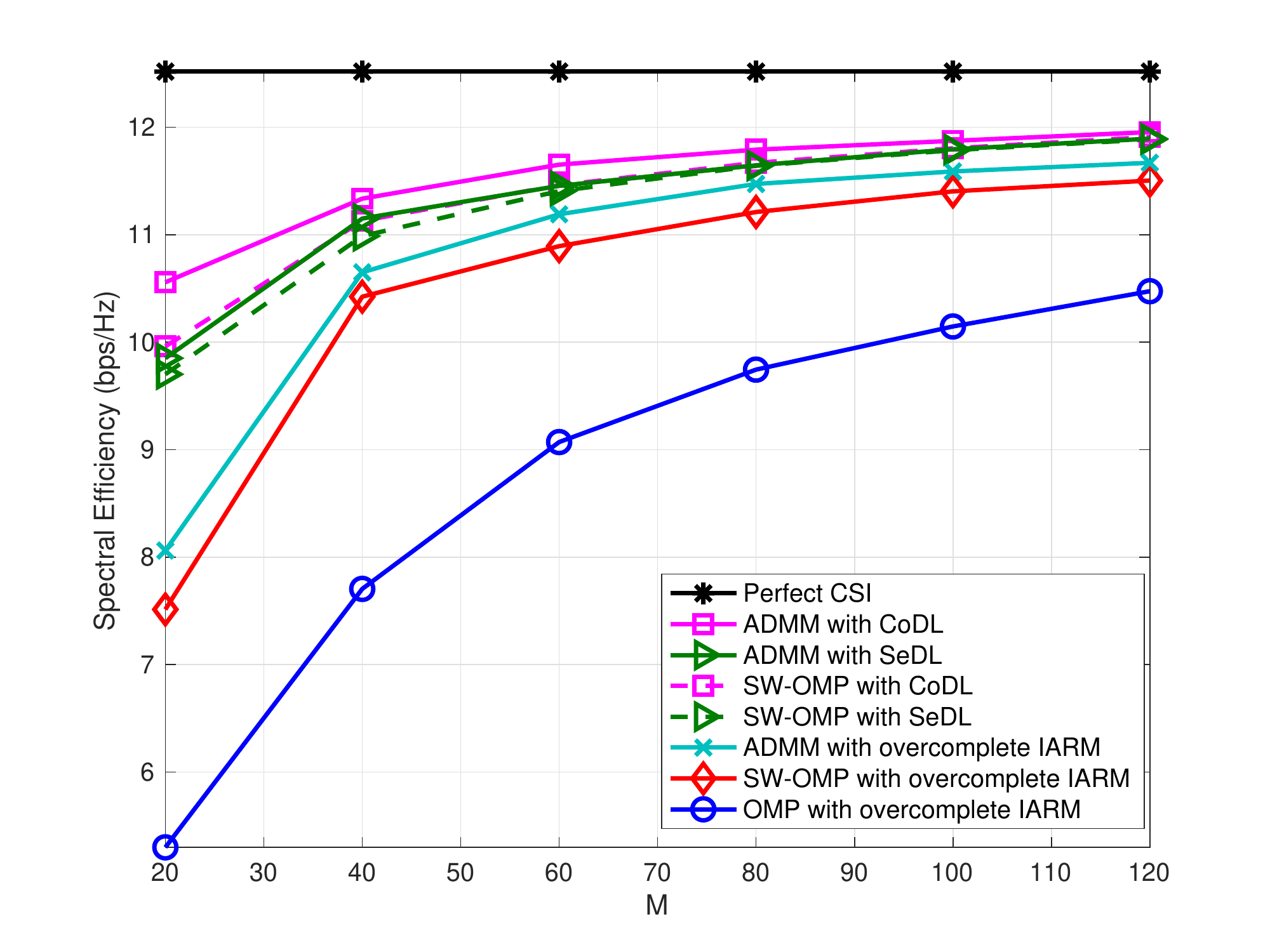}\label{fig:SE_M}}
	\caption{Comparisons of NMSE and SE performances for ULA versus number of training OFDM symbols using various sparse coding algorithms and sparsifying dictionaries. } \label{fig:MSE_SE_M} 
\end{figure} 
We then compare the channel estimation performances of various sparse coding algorithms and sparsifying dictionaries in \figref{fig:MSE_M} as a function of the number of training OFDM symbols.
The performance metric for channel estimation is the normalized mean squared error (NMSE), defined as 
\begin{equation*} 
\text{NMSE}=\frac{1}{\Nc}\sum_{c=0}^{\Nc-1}\frac{\| \hbH[c]- \bH[c]\|_F^2}{\| \bH[c]\|_F^2}. 
\end{equation*} 
The sparse coding algorithms, including OMP, SW-OMP and ADMM are used for channel estimation with overcomplete IARM dictionary, while SW-OMP and ADMM-based sparse coding are also evaluated with the dictionaries learned by CoDL and SeDL for comparison.    
It can be seen that when the learned dictionaries from CoDL and SeDL are used, the NMSE performances of SW-OMP outperform all the cases based on an IARM dictionary.  
This corroborates the effectiveness of the learned dictionaries for sparse channel representation and training overhead reduction. Furthermore, the ADMM-based sparse coding together with the optimized dictionaries are able to further reduce the training overhead, compared to all other sparse coding algorithms and dictionaries.
Last but not least, though there is a performance gap between SeDL and CoDL, SeDL is of lower computational complexity. In other words, SeDL provides a better trade-off between complexity and channel estimation performances.

Following the evaluation of NMSE performance in \figref{fig:MSE_M}, the comparison of spectral efficiency (SE) performances under the same assumptions is then given in \figref{fig:SE_M}, as a function of the number of training OFDM symbols. The SE is computed by assuming fully-digital precoding and combining using estimates for the $\Ns$ dominant left and right singular vectors of the channel estimates \cite{RodGonVen:Frequency-domain-Compressive-Channel:18}. Specifically, the effective channels are defined as $ \bH_{\text{eff}}[c]\teq \big[\hbU_1[c]\big]^*_{:,1:\Ns} \bH[c] \big[\hbU_2[c]\big]_{:,1:\Ns}, \forall c\in\cI(\Nc)$, where $\hbU_1[c]$ and $\hbU_2[c]$ are the left and right singular vectors of the channel estimates $ \hbH[c]$. Then SE is defined as
\begin{align*}
	\text{SE}=\frac{1}{\Nc}\sum_{c=0}^{\Nc-1}\sum_{n=1}^{\Ns}\log\left(1+\frac{\text{SNR}}{\Ns}\lambda_n( \bH_{\text{eff}}[c])^2\right),
\end{align*}
where $\lambda_n( \bH_{\text{eff}}[c]), n=1,\ldots,\Ns$ are the singular values of each effective channel $ \bH_{\text{eff}}[c]$.
For comparison, SE with perfect CSI is also provided as an upper bound. 
As can be seen from \figref{fig:SE_M}, SE is significantly increased by using SW-OMP with the learned dictionaries compared to those of SW-OMP with overcomplete IARM dictionary. Similar results can be found for ADMM-based sparse coding algorithm. These have illustrated the effectiveness of the proposed DL and channel estimation algorithms from a more practical viewpoint of interest. 

\begin{figure}[!ht] 
	\centering
	\subfigure[Comparison of NMSE performances for UCA.]{
		\includegraphics[width=0.46\textwidth]{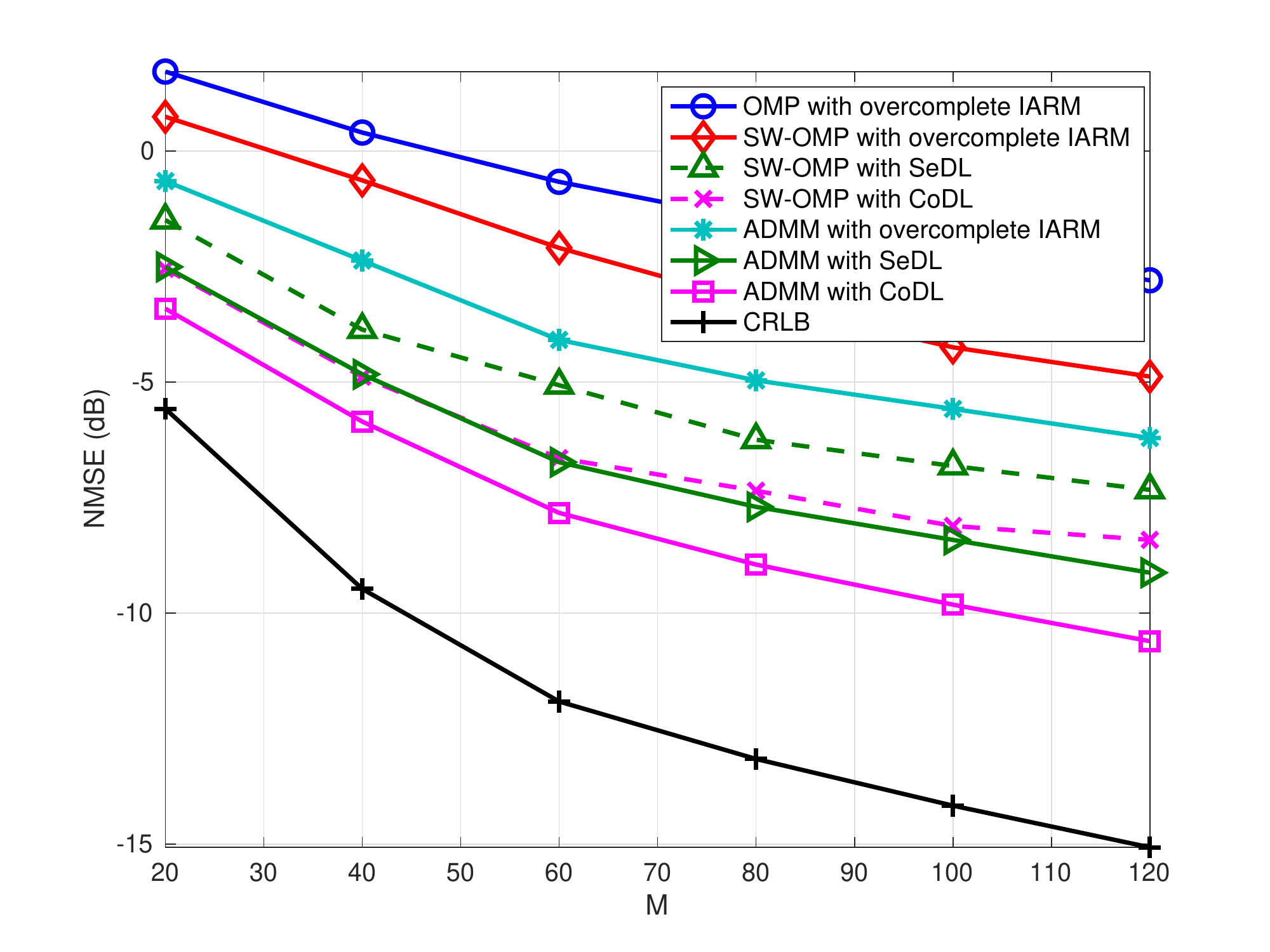}\label{fig:MSE_M_UCA}}
	\quad 
	\subfigure[Comparison of SE performances for UCA.]{
		\includegraphics[width=0.46\textwidth]{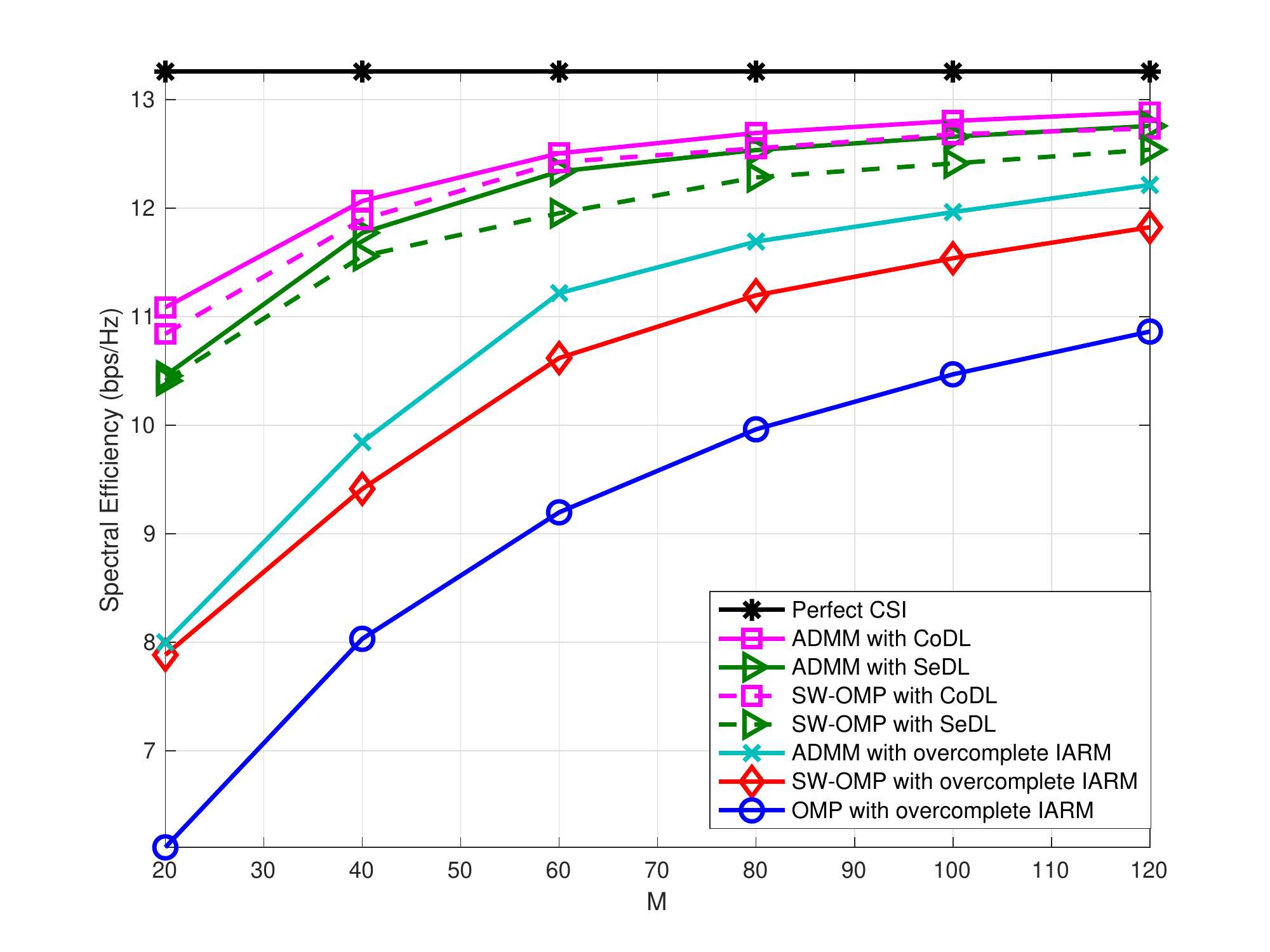}\label{fig:SE_M_UCA}}
	\caption{Comparisons of NMSE and SE performances for UCA versus number of training OFDM symbols using various sparse coding algorithms and sparsifying dictionaries. } \label{fig:MSE_SE_M_UCA} 
\end{figure} 
To evaluate the effectiveness of the proposed DL algorithms on different array geometries, the same procedures in \figref{fig:MSE_SE_M} have been evaluated for the uniform circular arrays (UCAs), where the presumed spacing distances between adjacent antennas are half-wavelength (e.g., see \cite{Du:Pattern-analysis-of-uniform:04}). Moreover, similar antenna gain/phase errors, mutual coupling \cite{EbeEscBie:Investigations-on-antenna-array:16} and antenna spacing disturbances as in \figref{fig:MSE_SE_M} are also included. From \figref{fig:MSE_M_UCA} and \ref{fig:SE_M_UCA}, it can be seen that the learned dictionaries from our proposed DL algorithms can result in significant performance gains for UCAs compared to the overcomplete IARM dictionary, both for SW-OMP or ADMM-based sparse coding algorithms. This is consistent with the case of ULAs and thus corroborates the applicability of the proposed DL algorithms for various (irregular) array geometries. 

 \begin{figure}[!ht]
	\centering 
	\subfigure[Comparison of NMSE performances.]{
		\includegraphics[width=0.46\textwidth]{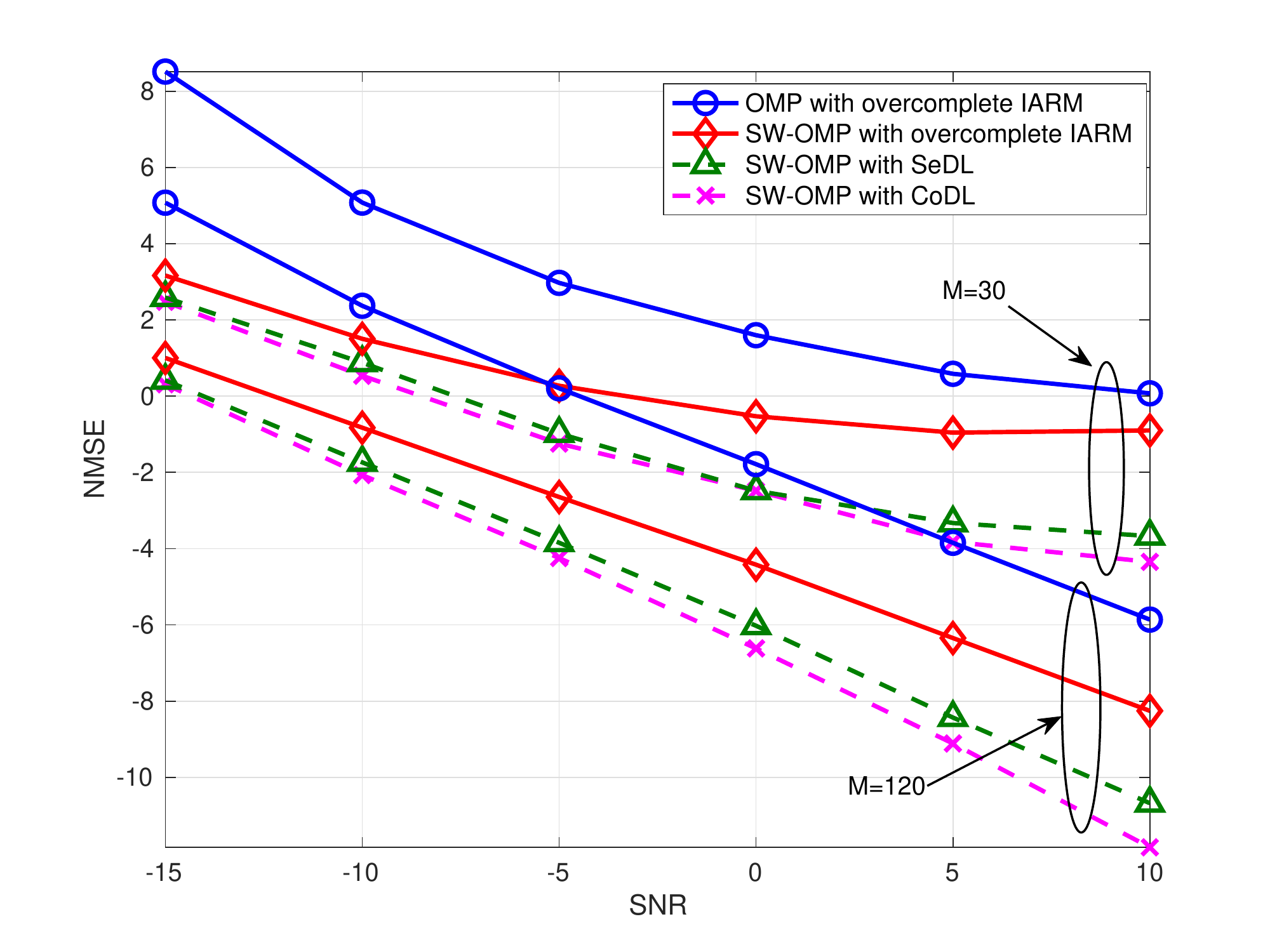}\label{fig:MSE_SNR}}
	\quad 
	\subfigure[Comparison of SE performances.]{
		\includegraphics[width=0.46\textwidth]{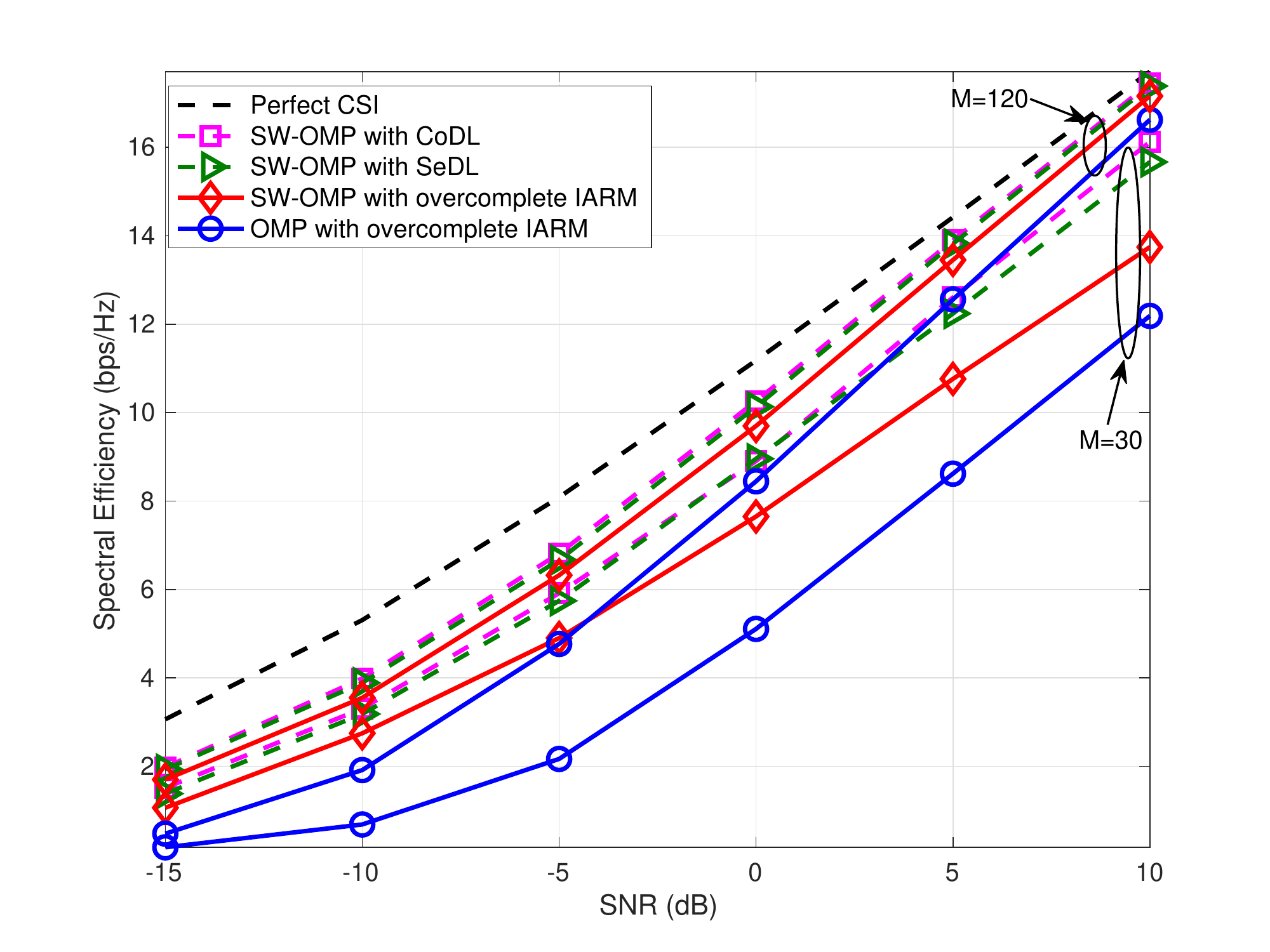}\label{fig:SE_SNR}}
	\caption{Comparisons of NMSE and SE performances of OMP and SW-OMP as a function of SNR using various sparsifying dictionaries. } \label{fig:MSE_SE_SNR} 
\end{figure} 

In \figref{fig:MSE_SE_SNR}, we compare the NMSE and SE performances of OMP and SW-OMP as a function of SNR, using overcomplete IARM dictionary or the learned dictionaries, and also include the impact of the number of training OFDM symbols. The parameters involved in this simulation are the same as \figref{fig:MSE_SE_M}. As can be seen, both the NMSE and SE performances of SW-OMP with learned dictionaries by CoDL and SeDL are superior to those using overcomplete IARM dictionary, even at a relatively low SNR. Of course, the performance gains of learned dictionaries at low SNR is relatively smaller, as the high noise level may mask the structural information of the channel so that the accuracy of learned dictionaries is deteriorated. Furthermore, the performance gains of the learned dictionaries are obvious when the number of training OFDM symbols is small, which means the DL methods can help to reduce the training overhead greatly.

\section{Conclusion} \label{sec:conclusions}

In this paper, we proposed two DL algorithms, i.e., CoDL and SeDL, to find the best sparsifying dictionaries for channel sparse representation in hybrid wideband mmWave MIMO systems. 
The CoDL and SeDL problems were formulated by exploiting the common sparsity properties within the large bandwidth. The
CoDL focused on the best combined dictionary,
while SeDL incorporated the detailed constraints on separable transmit and receive dictionaries, and thus SeDL can achieve a better trade-off between performance gains and computational complexity.
It has been shown that the learned dictionaries from both CoDL and SeDL can result in a sparser channel representation, compared to commonly adopted overcomplete Fourier dictionary, especially when there exist irregular array geometries and array uncertainties. 
With the learned dictionaries, various compressive channel estimation techniques can be applied for CSI acquisition at low SNR with much reduced training overhead, which has been corroborated via numerical simulations with different system configurations, array geometries and channel environments.

\section*{Appendix} 

\section*{Calculation of FIM $\bI(\bxi)$}

In this section, we provide the detailed derivations for the expressions of FIM $\bI(\bxi)$ based on \eqref{equ:FIM}.
We first consider the diagonal blocks of the FIM  $\bI(\bxi)$ with respect to (w.r.t.) each type of above unknown parameter.
Specifically, the elements of the FIM $\bI(\bxi)$ w.r.t. the AODs/AOAs $(\forall i,j\in\cI(\Np\Nray))$ are given as

\begin{align} 
\left[\bI(\bxi)\right]_{\theta_i,\phi_j} &= \frac{2}{\sigma^2}\Re\left\{\frac{\partial \bmuw^*(\bxi)}{\partial \theta_i}\frac{\partial \bmuw(\bxi)}{\partial \phi_j}\right\} 
= \frac{2}{\sigma^2}\sum_{c=0}^{\Nc-1}\Re\Big\{\overline{g_{i}[c]}\frac{\partial \baT(\theta_i)^T\kron \baR(\phi_i)^*}{\partial \theta_i}\notag \\
& \kern 10pt \cdot\big(\olbGammaT\kron \bGammaR \big)^*\big(\olbCT\kron \bCR\big)^*\bPhiw^*  
\bPhiw \big(\olbCT\kron \bCR\big)\big(\olbGammaT\kron \bGammaR \big) \frac{\partial \overline{\baT}(\theta_j)\kron \baR(\phi_j)}{\partial \phi_j}g_{j}[c]\Big\} \notag \\
&= \frac{2}{\sigma^2}\sum_{c=0}^{\Nc-1}\Re\Big\{\tr\Big\{\overline{g_{i}[c]}g_{j}[c]\big(\olbGammaT\kron \bGammaR \big)^*\big(\olbCT\kron \bCR\big)^*\bPhiw^* \bPhiw \big(\olbCT\kron \bCR\big)\big(\olbGammaT\kron \bGammaR \big) \notag\\ 
& \kern 10 pt \cdot ~ \overline{\baT}(\theta_j)\frac{\partial {\baT(\theta_i)}^T }{\partial \theta_i}\kron \frac{\partial \baR(\phi_j) }{\partial \phi_j} \baR(\phi_i)^*\Big\}\Big\}, 
\end{align} 
where, recalling the expression of $\baR(\phi_j)$ in \eqref{equ:aR}, the derivative of $\baR(\phi_j)$ w.r.t. $\phi_j$ is 
\begin{align}
\left[\frac{\partial \baR(\phi_j)}{\partial \phi_j}\right]_{m} = -\j\frac{2\pi(md+\epsilon_{\r,m})}{ \lambda}\cos(\phi_j) \left[  \baR(\phi_j) \right]_{m}, \forall m=0,\ldots,\Nr-1,
\end{align} 
and the derivative of $\baT(\theta_i)$ w.r.t. $\theta_i$ can be obtained accordingly. Likewise, we have

\vspace*{-3mm}
\begin{align}
\left[\bI(\bxi)\right]_{\phi_i,\theta_j} &= \frac{2}{\sigma^2}\sum_{c=0}^{\Nc-1}\Re\left\{\tr\left\{\overline{g_{i}[c]}g_{j}[c]\big(\olbGammaT\kron \bGammaR \big)^*\big(\olbCT\kron \bCR\big)^*\bPhiw^* \bPhiw \big(\olbCT\kron \bCR\big)\big(\olbGammaT\kron \bGammaR \big)\right.\right.\notag\\ 
&\left.\left. \kern 10 pt \cdot ~ \frac{\partial \overline{\baT}(\theta_j) }{\partial \theta_j}{\baT(\theta_i)}^T\kron \baR(\phi_j)\frac{\partial \baR(\phi_i)^* }{\partial \phi_i} \right\}\right\},
\end{align} 
\begin{align}
[\bI(\bxi)]_{\theta_i,\theta_j} 
& =  \frac{2}{\sigma^2} \sum_{c=0}^{\Nc-1}\Re\left\{\tr\left\{\overline{g_{i}[c]}g_{j}[c]\big(\olbGammaT\kron \bGammaR \big)^*\big(\olbCT\kron \bCR\big)^*\bPhiw^* \bPhiw \big(\olbCT\kron \bCR\big)\big(\olbGammaT\kron \bGammaR \big)\right.\right.\notag\\ 
&\left.\left. \kern 10 pt \cdot \frac{\partial \overline{\baT}(\theta_j)}{\partial \theta_j}\frac{ \partial \baT(\theta_i)^T}{\partial \theta_i}\kron\baR(\phi_j) \baR(\phi_i)^*\right\}\right\},
\end{align}
\begin{align}
\left[\bI(\bxi)\right]_{\phi_i,\phi_j} &= \frac{2}{\sigma^2}\sum_{c=0}^{\Nc-1}\Re\left\{\tr\left\{\overline{g_{i}[c]}g_{j}[c]\big(\olbGammaT\kron \bGammaR \big)^*\big(\olbCT\kron \bCR\big)^*\bPhiw^* \bPhiw \big(\olbCT\kron \bCR\big)\big(\olbGammaT\kron \bGammaR \big)\right.\right.\notag\\ 
&\left.\left. \kern 10 pt \cdot ~\overline{\baT}(\theta_j)\baT(\theta_i)^T\kron \frac{\partial \baR(\phi_j)}{\partial \phi_j} \frac{\partial \baR(\phi_i)^*}{\partial \phi_i}\right\}\right\}.
\end{align}
Similar computations for the elements of the FIM $\bI(\bxi)$ w.r.t. the path gains $(\forall i,j\in\cI(\Np\Nray),\forall c \in \cI(\Nc))$ yield
\begin{align} 
[\bI(\bxi)]_{g^\R_{i}[c],g^\R_{j}[c]} &= \frac{2}{\sigma^2} \Re\left\{ \frac{\partial \bmuw^*(\bxi)}{\partial g_i^\R[c]}  \frac{\partial \bmuw(\bxi)}{\partial g_j^\R[c]} \right\} 
= \frac{2}{\sigma^2}\Re \left\{\tr\left\{ \big(\olbGammaT\kron \bGammaR \big)^*\big(\olbCT\kron \bCR\big)^*\bPhiw^* \right.\right.\notag\\
&\kern 10pt \left.\left. \cdot\bPhiw \big(\olbCT\kron \bCR\big)\big(\olbGammaT\kron \bGammaR \big)  
\big(\overline{\baT}(\theta_j)\baT(\theta_i)^T\big) \kron \big(\baR(\phi_j)\baR(\phi_i)^*\big)   \right\}   \right\}, 
\end{align}
and $[\bI(\bxi)]_{g^\I_{i}[c],g^\I_{j}[c]} = [\bI(\bxi)]_{g^\R_{i}[c],g^\R_{j}[c]}$, and 
\begin{align}
[\bI(\bxi)]_{g^\R_{i}[c],g^\I_{j}[c]}
& = \frac{2}{\sigma^2}\Re \left\{\tr\left\{ \j\big(\olbGammaT\kron \bGammaR \big)^*\big(\olbCT\kron \bCR\big)^*\bPhiw^* \bPhiw \big(\olbCT\kron \bCR\big)\big(\olbGammaT\kron \bGammaR \big) \right.\right.\notag\\
&\kern 10pt \left.\left. \cdot ~\overline{\baT}(\theta_j)\baT(\theta_i)^T \kron \baR(\phi_j)\baR(\phi_i)^*   \right\}   \right\}, 
\end{align} 
\begin{align}
[\bI(\bxi)]_{g^\I_{i}[c],g^\R_{j}[c]} 
& = \frac{2}{\sigma^2}\Re \left\{\tr\left\{ -\j \big(\olbGammaT\kron \bGammaR \big)^*\big(\olbCT\kron \bCR\big)^*\bPhiw^* \bPhiw \big(\olbCT\kron \bCR\big)\big(\olbGammaT\kron \bGammaR \big) \right.\right.\notag\\
&\kern 10pt \left.\left. \cdot \overline{\baT}(\theta_j)\baT(\theta_i)^T \kron \baR(\phi_j)\baR(\phi_i)^*   \right\}   \right\}.
\end{align} 
Similar computations for the elements of FIM $\bI(\bxi)$ w.r.t. the antenna gain errors yield
\begin{align}
[\bI(\bxi)]_{g_{\t,i},g_{\r,j}} & = \frac{2}{\sigma^2}  \Re\left\{ \frac{\partial \bmuw^*(\bxi)}{\partial g_{\t,i}}  \frac{\partial \bmuw(\bxi)}{\partial g_{\r,j}} \right\}\notag \\ 
& =  \frac{2}{\sigma^2} \sum_{c=0}^{\Nc-1} \Re\left\{ \bg^*[c] (\olbAT\star \bAR)^* (\diag\{e^{\j\phi_{\t,i}} \bee_i\}\kron \bGammaR^*)(\olbCT\kron\bCR)^*\bPhiw^*\bPhiw (\olbCT\kron\bCR)\right. \notag\\ 
& \kern 10pt \left. \cdot (\olbGammaT\kron\diag\{e^{\j\phi_{\r,j}}\bee_j\})(\olbAT\star \bAR) \bg[c]     \right\}, \quad \forall i\in\cI(\Nt),j\in\cI(\Nr),
\end{align} 
where $\bee_i$ is the $i$-th column of the identity matrix of appropriate dimension.  Besides, we have 
\begin{align}
[\bI(\bxi)]_{g_{\r,i},g_{\t,j}}  
& =  \frac{2}{\sigma^2} \sum_{c=0}^{\Nc-1} \Re\left\{ \bg^*[c] (\olbAT\star \bAR)^* (\bGammaT^T\kron\diag\{e^{-\j\phi_{\r,i}}\bee_i\})(\olbCT\kron\bCR)^*\bPhiw^*\bPhiw (\olbCT\kron\bCR)\right. \notag\\ 
& \kern 10pt \left. \cdot (\diag\{e^{-\j\phi_{\t,j}} \bee_j\}\kron \bGammaR)(\olbAT\star \bAR) \bg[c]     \right\}, \quad \forall i\in\cI(\Nr),j\in\cI(\Nt),
\end{align} 
\begin{align}
[\bI(\bxi)]_{g_{\t,i},g_{\t,j}} 
& =  \frac{2}{\sigma^2} \sum_{c=0}^{\Nc-1} \Re\left\{ \bg^*[c] (\olbAT\star \bAR)^* (\diag\{e^{\j\phi_{\t,i}} \bee_i\}\kron \bGammaR^*)(\olbCT\kron\bCR)^*\bPhiw^*\bPhiw (\olbCT\kron\bCR)\right. \notag\\
& \kern 10pt \left. \cdot(\diag\{e^{-\j\phi_{\t,j}} \bee_j\}\kron \bGammaR) (\olbAT\star \bAR) \bg[c]  \right\}, \quad \forall i,j\in\cI(\Nt),
\end{align} 
\begin{align}
[\bI(\bxi)]_{g_{\r,i},g_{\r,j}} 
& =  \frac{2}{\sigma^2} \sum_{c=0}^{\Nc-1} \Re\left\{ \bg^*[c] (\olbAT\star \bAR)^*  (\bGammaT^T\kron\diag\{e^{-\j\phi_{\r,i}}\bee_i\})(\olbCT\kron\bCR)^*\bPhiw^*\bPhiw (\olbCT\kron\bCR)\right. \notag\\
& \kern 10pt \left. \cdot(\olbGammaT\kron\diag\{e^{\j\phi_{\r,j}}\bee_j\}) (\olbAT\star \bAR) \bg[c]     \right\} , \quad \forall i,j\in\cI(\Nr).
\end{align}
Similar computations for the elements of FIM $\bI(\bxi)$ w.r.t. the antenna phase errors yield
\begin{align}
[\bI(\bxi)]_{\phi_{\t,i},\phi_{\t,j}} & = \frac{2}{\sigma^2}  \Re\left\{ \frac{\partial \bmuw^*(\bxi)}{\partial \phi_{\t,i}}  \frac{\partial \bmuw(\bxi)}{\partial \phi_{\t,j}} \right\}\notag \\ 
& =  \frac{2}{\sigma^2} \sum_{c=0}^{\Nc-1} \Re\left\{ \bg^*[c] (\olbAT\star \bAR)^*   (\diag\{\j g_{\t,i}e^{\j\phi_{\t,i}}\bee_i\}\kron \bGammaR^* ) (\olbCT\kron\bCR)^*\bPhiw^*\bPhiw (\olbCT\kron\bCR)\right. \notag\\
& \kern 10pt \left. \cdot  (\diag\{-\j g_{\t,j}e^{-\j\phi_{\t,j}}\bee_j\}\kron \bGammaR )  (\olbAT\star \bAR) \bg[c]     \right\}, \quad \forall i,j \in\cI(\Nt),
\end{align}
and 
\begin{align}
[\bI(\bxi)]_{\phi_{\r,i},\phi_{\r,j}}  
& =  \frac{2}{\sigma^2} \sum_{c=0}^{\Nc-1} \Re\left\{ \bg^*[c] (\olbAT\star \bAR)^*    (\bGammaT^T\kron \diag\{-\j g_{\r,i}e^{-\j\phi_{\r,i}}\bee_i\} ) (\olbCT\kron\bCR)^*\bPhiw^*\bPhiw (\olbCT\kron\bCR)\right. \notag\\
& \kern 10pt \left. \cdot (\olbGammaT\kron \diag\{\j g_{\r,j}e^{\j\phi_{\r,j}}\bee_j\}  )  (\olbAT\star \bAR) \bg[c]     \right\}, \quad \forall i,j \in\cI(\Nr),
\end{align}
\begin{align}
[\bI(\bxi)]_{\phi_{\t,i},\phi_{\r,j}} 
& =  \frac{2}{\sigma^2} \sum_{c=0}^{\Nc-1} \Re\left\{ \bg^*[c] (\olbAT\star \bAR)^*   (\diag\{\j g_{\t,i}e^{\j\phi_{\t,i}}\bee_i\}\kron \bGammaR^* ) (\olbCT\kron\bCR)^*\bPhiw^*\bPhiw (\olbCT\kron\bCR)\right. \notag\\
& \kern 10pt \left. \cdot  (\olbGammaT\kron \diag\{\j g_{\r,j}e^{\j\phi_{\r,j}}\bee_j\}  ) (\olbAT\star \bAR) \bg[c]     \right\}, \quad \forall i\in\cI(\Nt), j\in\cI(\Nr),
\end{align}
\begin{align}
[\bI(\bxi)]_{\phi_{\r,i},\phi_{\t,j}} 
& =  \frac{2}{\sigma^2} \sum_{c=0}^{\Nc-1} \Re\left\{ \bg^*[c] (\olbAT\star \bAR)^*    (\bGammaT^T\kron \diag\{-\j g_{\r,i}e^{-\j\phi_{\r,i}}\bee_i\} ) (\olbCT\kron\bCR)^*\bPhiw^*\bPhiw (\olbCT\kron\bCR)\right. \notag\\
& \kern 10pt \left. \cdot (\diag\{-\j g_{\t,j}e^{-\j\phi_{\t,j}}\bee_j\}\kron \bGammaR )  (\olbAT\star \bAR) \bg[c]     \right\}, \quad \forall i\in\cI(\Nr), j\in\cI(\Nt).
\end{align}
Similar computations for the elements of FIM $\bI(\bxi)$ w.r.t. the antenna spacing errors yield
\begin{align}
[\bI(\bxi)]_{\epsilon_{\t,i},\epsilon_{\r,j}} & = \frac{2}{\sigma^2}  \Re\left\{ \frac{\partial \bmuw^*(\bxi)}{\partial \epsilon_{\t,i}}  \frac{\partial \bmuw(\bxi)}{\partial \epsilon_{\r,j}} \right\}\notag \\  
& =  \frac{2}{\sigma^2} \sum_{c=0}^{\Nc-1} \Re\left\{  \bg^*[c]  \Big(\frac{\partial \olbAT}{\partial \epsilon_{\t,i}}\star \bAR\Big)^*   (\olbGammaT\kron\bGammaR)^* (\olbCT\kron\bCR)^*\bPhiw^*\bPhiw (\olbCT\kron\bCR)\right. \notag\\
& \kern 10pt \left. \cdot (\olbGammaT\kron\bGammaR)  \Big(\olbAT\star \frac{\partial \bAR}{\partial \epsilon_{\r,j}}\Big)  \bg[c]     \right\},\quad \forall i\in\cI(\Nt), j\in\cI(\Nr),
\end{align}
where the derivative of $\bAR$ w.r.t. $\epsilon_{\r,j}$ is expressed as
\begin{align}
\left[\frac{\partial \bAR}{\partial \epsilon_{\r,j}}\right]_{:,\ell} = \frac{\partial \baR(\phil)}{\partial \epsilon_{\r,j}} = -\j\frac{2\pi}{\lambda}\sin(\phil)[\baR(\phil)]_j\cdot\bee_j,\quad \forall \ell\in\cI(\Np\Nray),j\in\cI(\Nr),
\end{align}
and the derivative of $\bAT$ w.r.t. $\epsilon_{\t,i}$ can be expressed accordingly. Besides, we have
\begin{align} 
[\bI(\bxi)]_{\epsilon_{\r,i},\epsilon_{\t,j}} 
& =  \frac{2}{\sigma^2} \sum_{c=0}^{\Nc-1} \Re\left\{  \bg^*[c]  \Big( \olbAT\star \frac{\partial\bAR}{\partial \epsilon_{\r,i}}\Big)^*  (\olbGammaT\kron\bGammaR)^* (\olbCT\kron\bCR)^*\bPhiw^*\bPhiw (\olbCT\kron\bCR)\right. \notag\\
& \kern 10pt \left. \cdot (\olbGammaT\kron\bGammaR)  \Big(\frac{\partial \olbAT}{\partial \epsilon_{\t,j}}\star \bAR\Big) \bg[c] \right\},\quad \forall i\in\cI(\Nr), j\in\cI(\Nt),
\end{align} 
\begin{align} 
[\bI(\bxi)]_{\epsilon_{\t,i},\epsilon_{\t,j}} 
& =  \frac{2}{\sigma^2} \sum_{c=0}^{\Nc-1} \Re\left\{  \bg^*[c]  \Big(\frac{\partial \olbAT}{\partial \epsilon_{\t,i}}\star \bAR\Big)^*  (\olbGammaT\kron\bGammaR)^* (\olbCT\kron\bCR)^*\bPhiw^*\bPhiw (\olbCT\kron\bCR)\right. \notag\\
& \kern 10pt \left. \cdot (\olbGammaT\kron\bGammaR)  \Big(\frac{\partial \olbAT}{\partial \epsilon_{\t,j}}\star \bAR\Big) \bg[c] \right\}, \quad \forall i,j\in\cI(\Nt),
\end{align} 
\begin{align}
[\bI(\bxi)]_{\epsilon_{\r,i},\epsilon_{\r,j}}  
& =  \frac{2}{\sigma^2} \sum_{c=0}^{\Nc-1} \Re\left\{  \bg^*[c]  \Big( \olbAT\star \frac{\partial\bAR}{\partial \epsilon_{\r,i}}\Big)^*  (\olbGammaT\kron\bGammaR)^* (\olbCT\kron\bCR)^*\bPhiw^*\bPhiw (\olbCT\kron\bCR)\right. \notag\\
& \kern 10pt \left. \cdot (\olbGammaT\kron\bGammaR) \Big(\olbAT\star \frac{\partial \bAR}{\partial \epsilon_{\r,j}}\Big)  \bg[c]     \right\},\quad \forall i,j \in\cI(\Nr).
\end{align} 
Finally, similar computations for the elements of FIM $\bI(\bxi)$ w.r.t. the coupling matrix coefficients yield
\begin{align}
[\bI(\bxi)]_{c_{\t,i,j},c_{\r,m,n}}  & = \frac{2}{\sigma^2}  \Re\left\{ \frac{\partial \bmuw^*(\bxi)}{\partial c_{\t,i,j}}  \frac{\partial \bmuw(\bxi)}{\partial c_{\r,m,n}} \right\}\notag \\  
& =  \frac{2}{\sigma^2} \sum_{c=0}^{\Nc-1} \Re\left\{  \bg^*[c]  ( \olbAT \star \bAR)^*   (\olbGammaT\kron\bGammaR)^* \Big(\frac{\partial \olbCT}{\partial c_{\t,i,j}}\kron\bCR \Big)^*\bPhiw^*\bPhiw \Big(\olbCT\kron\frac{\partial \bCR}{\partial c_{\r,m,n}}\Big)\right. \notag\\ 
& \kern 10pt \left. \cdot (\olbGammaT\kron\bGammaR)  (\olbAT\star  \bAR)  \bg[c]     \right\},\quad \forall 1<i<j<\Nt, 1<m<n<\Nr,
\end{align}
where the derivative of $\bCR$ w.r.t. $c_{\r,m,n}$ is an $\Nr\times\Nr$ matrix with ones at the indices of $c_{\r,m,n}$ and zeros otherwise, and the derivative of $\bCT$ w.r.t. $c_{\t,i,j}$ can be expressed accordingly. Besides, we have 
\begin{align}
[\bI(\bxi)]_{c_{\r,i,j},c_{\t,m,n}}     
& =  \frac{2}{\sigma^2} \sum_{c=0}^{\Nc-1} \Re\left\{  \bg^*[c]  ( \olbAT \star \bAR)^*   (\olbGammaT\kron\bGammaR)^* \Big(\olbCT\kron \frac{\partial \bCR}{\partial c_{\r,i,j}} \Big)^*\bPhiw^*\bPhiw \Big(\frac{\partial\olbCT}{\partial c_{\t,m,n}}\kron \bCR\Big)\right. \notag\\ 
& \kern 10pt \left. \cdot (\olbGammaT\kron\bGammaR)  (\olbAT\star  \bAR)  \bg[c]     \right\},\quad \forall 1<i<j<\Nr, 1<m<n<\Nt,
\end{align}
\begin{align}
[\bI(\bxi)]_{c_{\t,i,j},c_{\t,m,n}}   
& =  \frac{2}{\sigma^2} \sum_{c=0}^{\Nc-1} \Re\left\{  \bg^*[c]  ( \olbAT \star \bAR)^*   (\olbGammaT\kron\bGammaR)^* \Big(\frac{\partial \olbCT}{\partial c_{\t,i,j}}\kron\bCR \Big)^*\bPhiw^*\bPhiw \Big(\frac{\partial\olbCT}{\partial c_{\t,m,n}}\kron \bCR\Big)\right. \notag\\ 
& \kern 10pt \left. \cdot (\olbGammaT\kron\bGammaR)  (\olbAT\star  \bAR)  \bg[c]     \right\},\quad \forall 1<i<j<\Nt, 1<m<n<\Nt,
\end{align}
\begin{align}
[\bI(\bxi)]_{c_{\r,i,j},c_{\r,m,n}}   
& =  \frac{2}{\sigma^2} \sum_{c=0}^{\Nc-1} \Re\left\{  \bg^*[c]  ( \olbAT \star \bAR)^*   (\olbGammaT\kron\bGammaR)^* \Big(\olbCT\kron \frac{\partial \bCR}{\partial c_{\r,i,j}} \Big)^*\bPhiw^*\bPhiw \Big(\olbCT\kron\frac{\partial \bCR}{\partial c_{\r,m,n}}\Big)\right. \notag\\ 
& \kern 10pt \left. \cdot (\olbGammaT\kron\bGammaR)  (\olbAT\star  \bAR)  \bg[c]     \right\},\quad \forall 1<i<j<\Nr, 1<m<n<\Nr.
\end{align}
Note that the off-diagonal blocks of the FIM  $\bI(\bxi)$ between two different types of parameters can be obtained similarly following \eqref{equ:FIM} and the derivations of above diagonal blocks, and thus are omitted for space limitation.
The complete FIM $\bI(\bxi)$ is obtained as follows 
\begin{align}\label{equ:FIMfinal}
\bI(\bxi) = \Bigg[
\begin{matrix}
\bI^{(1,1)}(\bxi) & \bI^{(1,2)}(\bxi) \\
\bI^{(2,1)}(\bxi) & \bI^{(2,2)}(\bxi)
\end{matrix} \Bigg],
\end{align}
in which the sub-matrix $\bI^{(1,1)}(\bxi)$ is defined as 
\begin{align} 
\bI^{(1,1)}(\bxi) \teq \left[
\begin{smallmatrix} 
[\bI(\bxi)]_{\btheta,\btheta} & [\bI(\bxi)]_{\btheta,\bphi} & [\bI(\bxi)]_{\btheta,\bg[0]} & \cdots &[\bI(\bxi)]_{\btheta,\bg[\Nc-1]}  \\
[\bI(\bxi)]_{\bphi,\btheta} & [\bI(\bxi)]_{\bphi,\bphi} & [\bI(\bxi)]_{\bphi,\bg[0]} & \cdots &[\bI(\bxi)]_{\bphi,\bg[\Nc-1]}  \\
[\bI(\bxi)]_{\bg[0],\btheta} & [\bI(\bxi)]_{\bg[0],\bphi} & [\bI(\bxi)]_{\bg[0],\bg[0]} & \cdots &[\bI(\bxi)]_{\bg[0],\bg[\Nc-1]}  \\
\vdots & \vdots & \vdots & \ddots &\vdots&  \\
[\bI(\bxi)]_{\bg[\Nc{-}1],\btheta} & [\bI(\bxi)]_{\bg[\Nc{-}1],\bphi} & [\bI(\bxi)]_{\bg[\Nc-1],\bg[0]} & \cdots &[\bI(\bxi)]_{\bg[\Nc-1],\bg[\Nc-1]}  \\	  
\end{smallmatrix} 
\right],
\end{align} 
containing the Fisher information between AOAs/AODs and channel gains. Similarly, $\bI^{(1,2)}(\bxi)$ is given as
\begin{align}
\bI^{(1,2)}(\bxi) \teq \left[
\begin{smallmatrix}
[\bI(\bxi)]_{\bm g_{\t},\btheta} & [\bI(\bxi)]_{\bm g_{\t},\bphi} & [\bI(\bxi)]_{\bm g_{\t},\bg[0]} & \cdots &[\bI(\bxi)]_{\bm g_{\t},\bg[\Nc-1]} \\	 
[\bI(\bxi)]_{\bm g_{\r},\btheta} & [\bI(\bxi)]_{\bm g_{\r},\bphi} & [\bI(\bxi)]_{\bm g_{\r},\bg[0]} & \cdots &[\bI(\bxi)]_{\bm g_{\r},\bg[\Nc-1]} \\
[\bI(\bxi)]_{\bm \phi_{\t},\btheta} & [\bI(\bxi)]_{\bm \phi_{\t},\bphi} & [\bI(\bxi)]_{\bm \phi_{\t},\bg[0]} & \cdots &[\bI(\bxi)]_{\bm \phi_{\t},\bg[\Nc-1]} \\
[\bI(\bxi)]_{\bm \phi_{\r},\btheta} & [\bI(\bxi)]_{\bm \phi_{\r},\bphi} & [\bI(\bxi)]_{\bm \phi_{\r},\bg[0]} & \cdots &[\bI(\bxi)]_{\bm \phi_{\r},\bg[\Nc-1]} \\
[\bI(\bxi)]_{\bm \epsilon_{\t},\btheta} & [\bI(\bxi)]_{\bm \epsilon_{\t},\bphi} & [\bI(\bxi)]_{\bm \epsilon_{\t},\bg[0]} & \cdots &[\bI(\bxi)]_{\bm \epsilon_{\t},\bg[\Nc-1]} \\
[\bI(\bxi)]_{\bm \epsilon_{\r},\btheta} & [\bI(\bxi)]_{\bm \epsilon_{\r},\bphi} & [\bI(\bxi)]_{\bm \epsilon_{\r},\bg[0]} & \cdots &[\bI(\bxi)]_{\bm \epsilon_{\r},\bg[\Nc-1]} \\
[\bI(\bxi)]_{\bm c_{\t},\btheta} & [\bI(\bxi)]_{\bm c_{\t},\bphi} & [\bI(\bxi)]_{\bm c_{\t},\bg[0]} & \cdots &[\bI(\bxi)]_{\bm c_{\t},\bg[\Nc-1]} \\
[\bI(\bxi)]_{\bm c_{\r},\btheta} & [\bI(\bxi)]_{\bm c_{\r},\bphi} & [\bI(\bxi)]_{\bm c_{\r},\bg[0]} & \cdots &[\bI(\bxi)]_{\bm c_{\r},\bg[\Nc-1]}  
\end{smallmatrix}
\right],
\end{align}
which gathers the Fisher information between AOAs/AODs, channel gains and the remaining disturbance parameters. Moreover, $\bI^{(2,1)}(\bxi) = \big[\bI^{(1,2)}(\bxi)\big]^T$. Lastly, $\bI^{(2,2)}(\bxi)$ contains the Fisher information between remaining disturbance parameters and is expressed as 
\begin{align}
\bI^{(2,2)}(\bxi) \teq \left[
\begin{smallmatrix} 
[\bI(\bxi)]_{\bm g_{\t},\bm g_{\t}} & [\bI(\bxi)]_{\bm g_{\t},\bm g_{\r}}
&[\bI(\bxi)]_{\bm g_{\t},\bm \phi_{\t}}&[\bI(\bxi)]_{\bm g_{\t},\bm \phi_{\r}} & [\bI(\bxi)]_{\bm g_{\t},\bm \epsilon_{\t}} & [\bI(\bxi)]_{\bm g_{\t},\bm \epsilon_{\r}} & [\bI(\bxi)]_{\bm g_{\t},\bm c_{\t}} & [\bI(\bxi)]_{\bm g_{\t},\bm c_{\r}}\\	 
[\bI(\bxi)]_{\bm g_{\r},\bm g_{\t}} & [\bI(\bxi)]_{\bm g_{\r},\bm g_{\r}}
&[\bI(\bxi)]_{\bm g_{\r},\bm \phi_{\t}}&[\bI(\bxi)]_{\bm g_{\r},\bm \phi_{\r}} & [\bI(\bxi)]_{\bm g_{\r},\bm \epsilon_{\t}} & [\bI(\bxi)]_{\bm g_{\r},\bm \epsilon_{\r}} & [\bI(\bxi)]_{\bm g_{\r},\bm c_{\t}} & [\bI(\bxi)]_{\bm g_{\r},\bm c_{\r}}\\
[\bI(\bxi)]_{\bm \phi_{\t},\bm g_{\t}} & [\bI(\bxi)]_{\bm \phi_{\t},\bm g_{\r}}
&[\bI(\bxi)]_{\bm \phi_{\t},\bm \phi_{\t}}&[\bI(\bxi)]_{\bm \phi_{\t},\bm \phi_{\r}} & [\bI(\bxi)]_{\bm \phi_{\t},\bm \epsilon_{\t}} & [\bI(\bxi)]_{\bm \phi_{\t},\bm \epsilon_{\r}} & [\bI(\bxi)]_{\bm \phi_{\t},\bm c_{\t}} & [\bI(\bxi)]_{\bm \phi_{\t},\bm c_{\r}}\\
[\bI(\bxi)]_{\bm \phi_{\r},\bm g_{\t}} & [\bI(\bxi)]_{\bm \phi_{\r},\bm g_{\r}}
&[\bI(\bxi)]_{\bm \phi_{\r},\bm \phi_{\t}}&[\bI(\bxi)]_{\bm \phi_{\r},\bm \phi_{\r}} & [\bI(\bxi)]_{\bm \phi_{\r},\bm \epsilon_{\t}} & [\bI(\bxi)]_{\bm \phi_{\r},\bm \epsilon_{\r}} & [\bI(\bxi)]_{\bm \phi_{\r},\bm c_{\t}} & [\bI(\bxi)]_{\bm \phi_{\r},\bm c_{\r}}\\
[\bI(\bxi)]_{\bm \epsilon_{\t},\bm g_{\t}} & [\bI(\bxi)]_{\bm \epsilon_{\t},\bm g_{\r}}
&[\bI(\bxi)]_{\bm \epsilon_{\t},\bm \phi_{\t}}&[\bI(\bxi)]_{\bm \epsilon_{\t},\bm \phi_{\r}} & [\bI(\bxi)]_{\bm \epsilon_{\t},\bm \epsilon_{\t}} & [\bI(\bxi)]_{\bm \epsilon_{\t},\bm \epsilon_{\r}} & [\bI(\bxi)]_{\bm \epsilon_{\t},\bm c_{\t}} & [\bI(\bxi)]_{\bm \epsilon_{\t},\bm c_{\r}}\\
[\bI(\bxi)]_{\bm \epsilon_{\r},\bm g_{\t}} & [\bI(\bxi)]_{\bm \epsilon_{\r},\bm g_{\r}}
&[\bI(\bxi)]_{\bm \epsilon_{\r},\bm \phi_{\t}}&[\bI(\bxi)]_{\bm \epsilon_{\r},\bm \phi_{\r}} & [\bI(\bxi)]_{\bm \epsilon_{\r},\bm \epsilon_{\t}} & [\bI(\bxi)]_{\bm \epsilon_{\r},\bm \epsilon_{\r}} & [\bI(\bxi)]_{\bm \epsilon_{\r},\bm c_{\t}} & [\bI(\bxi)]_{\bm \epsilon_{\r},\bm c_{\r}}\\
[\bI(\bxi)]_{\bm c_{\t},\bm g_{\t}} & [\bI(\bxi)]_{\bm c_{\t},\bm g_{\r}}
&[\bI(\bxi)]_{\bm c_{\t},\bm \phi_{\t}}&[\bI(\bxi)]_{\bm c_{\t},\bm \phi_{\r}} & [\bI(\bxi)]_{\bm c_{\t},\bm \epsilon_{\t}} & [\bI(\bxi)]_{\bm c_{\t},\bm \epsilon_{\r}} & [\bI(\bxi)]_{\bm c_{\t},\bm c_{\t}} & [\bI(\bxi)]_{\bm c_{\t},\bm c_{\r}}\\
[\bI(\bxi)]_{\bm c_{\r},\bm g_{\t}} & [\bI(\bxi)]_{\bm c_{\r},\bm g_{\r}}
&[\bI(\bxi)]_{\bm c_{\r},\bm \phi_{\t}}&[\bI(\bxi)]_{\bm c_{\r},\bm \phi_{\r}} & [\bI(\bxi)]_{\bm c_{\r},\bm \epsilon_{\t}} & [\bI(\bxi)]_{\bm c_{\r},\bm \epsilon_{\r}} & [\bI(\bxi)]_{\bm c_{\r},\bm c_{\t}} & [\bI(\bxi)]_{\bm c_{\r},\bm c_{\r}}  
\end{smallmatrix}
\right].
\end{align}

\nocite{*}
\bibliographystyle{IEEEtran}
\bibliography{DLrefs}

\end{document}